\documentclass[12pt]
{iopart} 
\usepackage{iopams,verbatim} 

\newcommand{\mysectionapp}[1]{\section{#1}}
\newcommand{\mysubsectionapp}[1]{\addtocounter{subsection}{1}\subsection*{\Alph{section}\arabic{subsection}. #1}} 

\newtheorem{theorem}{Theorem}
\newtheorem{lemma}{Lemma}
\newtheorem{corollary}{Corollary}
\newtheorem{remark}{Remark}
\newtheorem{example}{Example}

\newcommand{\proof}{\noindent{\em Proof:} \ }
\newcommand{\enproof}{\hfill $\Box$ \vspace*{1ex}}
\newcommand{\enlem}{\end{lemma}}
\newcommand{\closedef}{\hfill $\Diamond$ \end{definition}}
\newcommand{\enth}{\hfill $\Diamond$ \end{theorem}}
\newcommand{\encor}{\hfill $\Diamond$ \end{corollary}}
\newcommand{\enprop}{\hfill $\Diamond$ \end{proposition}}
\newcommand{\encond}{\hfill $\Diamond$ \end{condition}}
\newcommand{\exam}[1]{\begin{example}\label{ex:#1}}
\newcommand{\enexam}{\QED\end{example}}
\newcommand{\beremark}[1]{\begin{remark}\label{rmk:#1}}

\newcommand{\mymathbb}[1]{{\mathbb #1}} 
\newcommand{\mymathsf}[1]{{\mathsf{#1}}} 

\newcommand{\dmn}{d}

\newcommand{\cA}{{\cal A}}
\newcommand{\sA}{\mymathsf{A}}
\newcommand{\cB}{{\cal B}}

\newcommand{\bC}{\mymathbb{C}}

\newcommand{\cE}{{\cal E}}
\newcommand{\sE}{\mymathsf{E}}

\newcommand{\cG}{{\cal G}}
\newcommand{\sF}{\mymathsf{F}}
\newcommand{\myF}{{\mymathbb{F}_{\dmn}}} 

\newcommand{\cN}{{\cal N}}
\newcommand{\sS}{\mymathsf{S}}
\newcommand{\sT}{\mymathsf{T}}

\newcommand{\sH}{\mymathsf{H}} 
\newcommand{\cI}{{\cal I}}

\newcommand{\cL}{{\cal L}}
\newcommand{\sL}{\mymathsf{L}}
\newcommand{\cM}{{\cal M}}
\newcommand{\sM}{\mymathsf{M}}
\newcommand{\sN}{\mymathsf{N}}
\newcommand{\cP}{{\cal P}}
\newcommand{\sP}{\mymathsf{P}}
\newcommand{\sQ}{\mymathsf{Q}}
\newcommand{\cQ}{{\cal Q}}

\newcommand{\cR}{{\cal R}}
\newcommand{\cS}{{\cal S}}
\newcommand{\cT}{{\cal T}}
\newcommand{\cU}{{\cal U}}
\newcommand{\cV}{{\cal V}}
\newcommand{\sV}{\mymathsf{V}}
\newcommand{\cX}{{\cal X}}
\newcommand{\sX}{\mymathsf{X}}
\newcommand{\sY}{\mymathsf{Y}}
\newcommand{\sW}{\mymathsf{W}}
\newcommand{\sU}{\mymathsf{U}}
\newcommand{\sZ}{\mymathsf{Z}}
\newcommand{\cY}{{\cal Y}}

\newcommand{\wbX}[1]{{X}^{#1}} 
\newcommand{\wbZ}[1]{{Z}^{#1}}

\newcommand{\vep}{\varepsilon}
\renewcommand{\subset}{\subseteq}
\renewcommand{\tilde}{\widetilde}
\renewcommand{\hat}{\widehat}
\newcommand{\Bar}{\overline}  
\renewcommand{\bar}{\Bar}

\newcommand{\mbm}[1]{\mbox{\boldmath $#1$}}

\newcommand{\defeq}{\stackrel{\rm def}{=}}

\newcommand{\cmple}{^{\rm c}}
\newcommand{\prcmple}{'^{{\rm c}}}

\newcommand{\SINT}{\mymathbb{Z}}
\newcommand{\SNN}{\mymathbb{N}}

\newcommand{\Expe}{\mymathbb{E}}

\newcommand{\Prob}{{\rm Pr}}

\newcommand{\transp}{^{\rm T}}

\newcommand{\tnsr}{\otimes}
\newcommand{\trace}{{\rm Tr}}

\newcommand{\lag}{\langle}
\newcommand{\rag}{\rangle}
\newcommand{\crd}[1]{|#1|}
\newcommand{\bra}[1]{\lag #1 |}
\newcommand{\ket}[1]{| #1 \rag}

\newcommand{\chf}{{\bf 1}}
\newcommand{\dpr}[2]{#1 \cdot #2}

\newcommand{\Hch}{{\sH}}

\newcommand{\Hgen}{{\sH^{\tnsr n}}}
\newcommand{\Dgen}{{\dmn^n}}

\newcommand{\Bop}{\sL} 

\newcommand{\Fen}{F_{\rm e}}

\newcommand{\imu}{{\rm i}} 
\newcommand{\Ebe}{N}

\newcommand{\ketbe}[1]{\ket{#1}}
\newcommand{\phasebe}{\omega}
\newcommand{\Xbe}{X}
\newcommand{\Zbe}{Z}

\newcommand{\Aso}{\sA}  
\newcommand{\Acn}[1]{\sA_{#1}}  
\newcommand{\CPex}{\cV} 

\newcommand{\vara}{l} 
\newcommand{\varaa}{l'} 
\newcommand{\varaaa}{u} 
\newcommand{\varii}{l} 
 
\newcommand{\vari}{s} 
\newcommand{\varj}{t} 

\newcommand{\vare}{e}
\newcommand{\vars}{\xi} 
\newcommand{\tvara}{s}
\newcommand{\tvarb}{t}

\newcommand{\rvy}{\mymathsf{Y}}
\newcommand{\rvestA}{\rvy_{\rm A}}
\newcommand{\rvestB}{\rvy_{\rm B}}
\newcommand{\rlestA}{Y_{\rm A}}
\newcommand{\rlestB}{Y_{\rm B}}

\newcommand{\varn}{m} 

\newcommand{\spn}{\mymathsf{span}\,}

\newcommand{\Id}{\cI} 

\newcommand{\rvx}{\sX} 
\newcommand{\rvz}{\sZ} 
\newcommand{\rvv}{\sV}
\newcommand{\rvs}{\sS}
\newcommand{\rls}{s}
\newcommand{\rvt}{\sT}
\newcommand{\rve}{\sE}
\newcommand{\rvw}{\sW}
\newcommand{\rvu}{\sU}

\newcommand{\vxy}{x}

\newcommand{\hMsubnoarg}[1]{\sM_{#1}} 
\newcommand{\hMsub}[2]{\sM_{#1}({#2})} 
 
\newcommand{\hMsubbar}[2]{\bar{\sM}_{#1}({#2})} 
\newcommand{\Ccl}{C}
\newcommand{\kcl}{\kappa} 
\newcommand{\erx}{\delta_X} 
\newcommand{\erz}{\delta_Z} 
\newcommand{\css}[2]{\mymathsf{CSS}(#1,#2)}
\newcommand{\dbar}[1]{\bar{\bar{#1}}}
\newcommand{\Nwd}[1]{T_{#1}}
\newcommand{\Se}[2]{S_{#1 #2}}
\newcommand{\Senew}[2]{S_{#1, #2}}
\newcommand{\myFpower}[1]{\mymathbb{F}_{\dmn}^{#1}}
\newcommand{\myFpowerarg}[2]{\mymathbb{F}_{#1}^{#2}}

\newcommand{\Pev}{P_{\Eve}}
\newcommand{\Qpl}[1]{\sQ^+} 
\newcommand{\Qmi}[1]{\sQ^-}

\newcommand{\rva}{\mymathsf{a}} 
\newcommand{\rvb}{\mymathsf{b}} 
\newcommand{\rvc}{\mymathsf{c}} 
\newcommand{\rvaa}{\rva'}
\newcommand{\rvaaa}{\rva''}

\newcommand{\pa}{p_{\rm a}}
\newcommand{\pb}{p_{\rm b}}
\newcommand{\pc}{p_{\rm c}}
\newcommand{\pccmp}{(1-p_{\rm c})}
\newcommand{\Bp}[1]{B(#1)}

\newcommand{\Pa}{P_0}
\newcommand{\Pb}{P_1}
\newcommand{\Qa}{Q_0}
\newcommand{\Qb}{Q_1}

\newcommand{\rlf}[2]{\hat{#1}} 

\newcommand{\xa}{x}
\newcommand{\xb}{x'}

\newcommand{\mua}{\mu} 
\newcommand{\mub}{\mu'} 
\newcommand{\nua}{\nu} 
\newcommand{\nub}{\nu} 
\newcommand{\hmo}{h} 
\newcommand{\Hcndnoarg}{h_{\rm c}} 
\newcommand{\Hcndr}[2]{\Hcndnoarg(#1,#2)} 

\newcommand{\Hcndn}[2]{\Hcndnoarg(#1,#2)}
\newcommand{\zrv}{0^n}
\newcommand{\flp}[1]{#1^{\mymathsf{f}}} 

\newcommand{\Eve}{\cA}
\newcommand{\vma}{\lambda_{m}}
\newcommand{\vmb}{\lambda'_{m}}
\newcommand{\Estar}{E^*} 
\newcommand{\vp}{p}
\newcommand{\vpa}{p_0}
\newcommand{\vpb}{p_1}
\newcommand{\sqi}[1]{_{#1}}
\newcommand{\Csm}{\Ccl} 
\newcommand{\Cbg}{\Ccl^{\perp}} 
\newcommand{\Cgood}{C}
\newcommand{\Jgood}{\Gamma}
\newcommand{\vra}{r}

\newcommand{\Ksp}{J} 
\newcommand{\crJ}{\Gamma} 
\newcommand{\tlJ}{\tilde{\Gamma}}
\newcommand{\Jof}[1]{\prm(\Jgood)} 
\newcommand{\Kof}[1]{\Ksp(#1)} 

\newcommand{\tdP}{p}
\newcommand{\codesubs}[1]{\cQ_{#1}}
\newcommand{\tsp}{M}
\newcommand{\tsptwo}[2]{\tsp_{#1}(#2)}
\newcommand{\tspbarone}[1]{\bar{\tsp}_{#1}}
\newcommand{\prm}{\pi}

\newcommand{\cnt}[1]{L_{#1}}
\newcommand{\cpf}{\cP_n}
\newcommand{\wkb}{} 
\newcommand{\Evn}{\cV_n} 

\newcommand{\Ensperm}{\cS_n}
\newcommand{\rvd}{\mymathsf{d}}
\newcommand{\pxp}[1]{Q_{#1}}

\newcommand{\rveta}{\mbm{\eta}}
\newcommand{\rvetaA}{\rveta^{\rm A}}
\newcommand{\rvetaB}{\rveta^{\rm B}}

\newcommand{\rvcl}{\mymathsf{n}} 
\newcommand{\rlcl}{n}
\newcommand{\rvms}{\mbm{\mu}} 
\newcommand{\rlms}{\mu} 
\newcommand{\rvxi}{\mbm{\xi}} 
\newcommand{\rvzeta}{\mbm{\zeta}} 

\newcommand{\rvss}{\rvs'} 
\newcommand{\rvsss}{(\rvy_{\rm A},\rvy_{\rm B})} 
\newcommand{\rvssss}{\rvs'''}
\newcommand{\rlss}{(\rltsift,\rlms,\rlcl)} 
\newcommand{\rlsss}{(Y_{\rm A},Y_{\rm B})} 
\newcommand{\rlssss}{s'''}
\newcommand{\rvpi}{\mbm{\prm}}
\newcommand{\rvk}{\mymathsf{k}}
\newcommand{\rvcode}{\mymathsf{C'}} 

\newcommand{\Fcnd}[1]{F_{#1}}
\newcommand{\Bprv}[1]{\Bp{P_{#1}}}
\newcommand{\rvgn}{\sG}
\newcommand{\sG}{\mymathsf{G}}

\newcommand{\rvxicode}{\rvxi_{\rm code}}
\newcommand{\rvxiest}{\rvxi_{\rm est}}

\newcommand{\rvzetacode}{\rvzeta_{\rm code}}
\newcommand{\rvzetaest}{\rvzeta_{\rm est}}

\newcommand{\Pxncnd}{P_{\rvxicode|\rvk=k,\rvss=\rlss,\rvsss=\rlsss}}

\newcommand{\Kd}{(2\ln \dmn)} 
\newcommand{\rvtsift}{\rvt_{\rm sift}}
\newcommand{\rltsift}{T_{\rm sift}}
\newcommand{\rvrb}{\alpha}
\newcommand{\Ejoint}{E_{1}}

\begin{document} 

\title[Reliability of CSS 
Codes and Security of Quantum Key Distribution]
{Reliability of Calderbank-Shor-Steane Codes and Security of
Quantum Key Distribution}

\author{Mitsuru Hamada}
\address{Quantum Computation and Information Project, ERATO Program\\
     Japan Science and Technology Agency\\
      5-28-3, Hongo, Bunkyo-ku, Tokyo 113-0033, Japan}
\eads{mitsuru@ieee.org}

\date{May, 2003}



\begin{abstract}
After Mayers (1996, 2001) gave a proof of the security
of the Bennett-Brassard 1984 (BB84) quantum key distribution protocol,
Shor and Preskill (2000) made a remarkable observation that
a Calderbank-Shor-Steane (CSS) code had been implicitly used
in the BB84 protocol, and suggested its security
could be proved by bounding the fidelity, say $F_n$, of the 
incorporated CSS code of length $n$ in the form
$1-F_n \le \exp[-n E+o(n)]$ for some positive number $E$.
This work presents such a number
$E=E(R)$ as a function of the rate of codes $R$,
and a threshold $R_0$ such that $E(R)>0$ whenever $R < R_0$,
which is larger than the
achievable rate based on the Gilbert-Varshamov bound
that is essentially due to Shor and Preskill (2000).
The codes in the present work are robust against fluctuations
of channel parameters, which fact is needed to establish the security
rigorously and was not proved 
for rates above the Gilbert-Varshamov
rate before in the literature. 
As a byproduct, the security of a modified BB84 protocol
against any joint (coherent) attacks is proved quantitatively.
\end{abstract}

\maketitle

\section{Introduction \label{ss:intro}}

The security of quantum key distribution (QKD), the aim of which is to share
a random secret string of digits between two parties,
has been said to rest on the principle of quantum mechanics
since the time of its proposal~\cite{BennettBrassard84}.
However, proofs of the security 
against a reasonably wide class of attacks were obtained only recently
on the first QKD protocol, which uses Wiesner's idea of conjugate
coding~\cite{wiesner83} and is called 
the Bennett-Brassard 1984 (BB84) protocol~\cite{BennettBrassard84}.
Since a preliminary report on such a proof of the security 
of the scheme was given by Mayers~\cite{mayers96},
there have been considerable
efforts to refine, strengthen or support this result in the literature
(e.g., \cite{mayers01acm,biham99,ShorPreskill00,GottesmanPreskill01,GottesmanLLP02,TamakiKI03}).
Especially,
Shor and Preskill~\cite{ShorPreskill00} (see also \cite[Section~III]{GottesmanPreskill01}) 
made a remarkable observation that
a Calderbank-Shor-Steane (CSS) quantum code had been implicitly used
in the BB84 protocol, and suggested if the fidelity, say $F_n$, of 
the incorporated Calderbank-Shor-Steane code~\cite{CalderbankShor96,steane96a}
goes to unity exponentially as the code-length $n$ grows large,
viz., $1-F_n \le \exp[-n E +o(n)]$ for some positive number $E$,
then the security of the BB84 protocol will be ensured 
in the sense that the mutual information between the shared key
and the data obtained by the eavesdropper is less than
$\exp[-n E +o(n)]$.
However, no one seems to have given such an exponent $E$ 
for CSS codes explicitly in the literature.
Thus, this paper is concerned with the problem of finding
such an exponent $E(R)$ as an explicit function of the rate $R$ of CSS codes.

The proviso for the security proof in this paper is as follows:
In the main text, we assume that the possible eavesdropper tries to obtain data
by performing an identical measurement on each `particle'
(what is really meant is the $\dmn$-level quantum system carrying a digit
from $\{ 0, \dots, \dmn-1 \}$, which is
typically assumed to be the polarization of a photon, a two-level system);
the two legitimate participants of the protocol can communicate with each other
by means of a classical noiseless `public channel'
that may be susceptible to passive eavesdropping but is free of
tampering;
we adopt the formalism developed by Kraus and others to describe
measurements
(e.g., \cite{HolevoLN,kraus71,hellwig95,kraus,preskillLNbook}).
We assume the so-called individual-attack assumption as mentioned above 
in order to discuss trade-offs between the level of attacks 
(including noises)
and the allowed rates of transmission of the key; without such an assumption, 
the level of attacks (often called error rates) could not be properly defined
for this purpose.
After this tractable case is worked out,
the security of a modified BB84 protocol
against any joint (coherent) attacks is proved quantitatively
in \ref{app:joint}.

Among others, it is proved that
a code of `balanced weight spectrum', i.e.,
a code whose weight distribution is almost proportional to the
binomial coefficients (when $\dmn=2$) attains the desired fidelity bound.
This would show the direction to designers of codes for QKD.
The code is robust against
fluctuations of channel parameters,
which is needed to complete the proof of the security rigorously
for rates beyond the Gilbert-Varshamov one
even in the case of individual attacks.
The channel parameters have to be estimated by the participants of the
BB84 to assess the level of eavesdropping,
and the robustness is necessary because the estimated channel parameters
are not exactly equal to the true ones in general. 
The robustness issue will be resolved
by utilizing the idea of universal codes~\cite{CsiszarKoerner,CsiszarKoerner81a} in information theory. A universal code means one whose structure does not depend on the channel characteristics.  

The CSS codes form a class of symplectic
(stabilizer or additive) codes~\cite{crss97,crss98,gottesman96},
and there exists a simple class of CSS codes, in which a CSS code
is specified by a classical code, say $C'$,
satisfying some condition on orthogonality. 
If we are content with correcting the errors of Hamming weight 
up to $\delta n/2$, where $\delta n$ is the minimum distance 
of $C'$,
exponential convergence of fidelity
immediately follows from the Gilbert-Varshamov
bound for CSS codes~\cite{CalderbankShor96} and Sanov's
theorem (Section~\ref{ss:discussions}), which is 
central in large deviation theory~\cite{DemboZeitouni,CoverThomas}.
Nevertheless, this argument only ensures the security of the BB48 protocol
of code rate up to $1-2 \hmo(\erx+\erz)$, 
where 
$\hmo$ is the base-two binary entropy function,
$\erz$ is the raw bit error rate 
in transmitting a bit encoded into an eigenvector
$\ket{0}$ or $\ket{1}$ of a Pauli operator, say, $Z$,
and $\erx$ is that with a bit encoded into $\ket{0}\pm\ket{1}$.
Note that the argument of Shor and Preskill~\cite{ShorPreskill00} can easily
be modified
to establish the rate $1-2 \hmo(\erx+\erz)$ for individual attacks
(Section~\ref{ss:discussions}).
The aim of this paper includes to obtain, 
in a rigorous manner, 
the better achievable rate $1-2\hmo\big((\erx+\erz)/2\big)$.
This rate seems essentially the same as the one previously mentioned
in the literature~\cite{ShorPreskill00}, \cite[Eq.~(38)]{GottesmanPreskill01}, 
though these papers focused on other issues
and gave no details on their codes achieving this higher rate.

We remark that
in comparing this paper's bound with the previously claimed ones,
we should care about the meaning of `error rates'.
Namely, strictly speaking, we should distinguish the error rates in this paper
from the `error rates' in security proofs for joint attacks. 
Specifically,
our $\erx$ and $\erz$ are parameters of the channel that represents 
the eavesdropper's attack on each digit
whereas it is natural to define
the `error rates' for joint attacks 
as some fictional random variables which are associated with 
the much larger channel that represents 
a general joint attack; In \ref{app:joint} of this paper,
the `error rates' for joint attacks will appear as
$P_{\rvxi'}(1)$ and $P_{\rvzeta'}(1)$, 
where $P_{\rvxi'}$ [$P_{\rvzeta'}$] is the {\em type}, i.e.,
the empirical distribution of the `sifted' part, 
or an even smaller part, $\rvxi'$ [$\rvzeta'$]
of the sequence of random variables $\rvxi$ [$\rvzeta$].

Results on exponential convergence of the fidelity of quantum codes
(quantum error-correcting codes) have
already been obtained by the present author
with random coding, which is a proof technique of Shannon's, 
over general symplectic codes~\cite{hamada01e,hamada01g,hamada02c}. 
These previous results, however, ensure only the existence of 
reliable symplectic codes, and use of symplectic codes 
other than CSS codes in QKD seems to require a quantum computer
to implement~\cite{ShorPreskill00}.
Thus, this paper will provide a rigorous but elementary proof 
that the fidelity $F_n$ of some CSS codes of rate $R$ satisfies
$1-F_n \le \exp[-n E(R) +o(n)]$ for some function $E(R)$ such that
$E(R) > 0$ whenever 
$R<1-2\hmo\big((\erx+\erz)/2\big)$.

Using this bound and
Schumacher's argument~\cite{schumacher96}, 
which related channel codes with quantum cryptography,
we prove the security of 
the BB84 protocol.
The proof to be presented below 
is basically a refinement of Shor and Preskill's.
%
Whereas use of two-level systems is often assumed
when symplectic codes or the BB84 protocol
are discussed in the literature,
most notions and results easily extend to $\dmn$-level
systems with an arbitrary prime $\dmn$.
Moreover, maybe contrary to one's expectation,
our analysis in
the case where $\dmn \ge 3$ will turn out 
to be more tractable than in the case where $\dmn=2$
except for the part treating channel estimation, 
so that we will begin with the easier case where $\dmn \ge 3$.

We neither touch on more practical issues such as the one on 
difficulty in preparing a single photon or 
how to implement $\dmn$-level systems,
nor treat more elaborated models allowing basis-dependent attacks and so
on~\cite{GottesmanLLP02}. 

We remark that 
there has already been a proposal
to use {\em two-way}\/ entanglement distillation protocols for QKD
in order to increase the maximum tolerable error rate~\cite{GottesmanLo01},
whereas the security of the BB84 protocol to be treated in this paper relies on
simpler quantum error-correcting (CSS) codes, which can be viewed
as {\em one-way}\/ entanglement distillation protocols.
The former class is still based on CSS codes, 
and would deserve further investigations.
However, 
we will stay around the simple class of protocols in this paper
in order to resolve the issues mentioned above. 

Attainable fidelity of codes given in this paper 
may also be interesting from a viewpoint of quantum computing
since CSS codes are well-suited for fault-tolerant quantum computing~\cite{steane99,gottesmanPhD}.
Incidentally, 
the technique (permutation argument) in the existence proof of CSS codes in this paper can be incorporated into
those of \cite{hamada01e,hamada01g,hamada02c} 
to show that the fidelity bounds of \cite{hamada01e,hamada01g,hamada02c} can be attained by robust symplectic codes.

The paper is organized as follows.
In Section~\ref{sec:CSS}, the needed notation on CSS codes is fixed
and a brief review on this class of codes is given. 
In Section~\ref{sec:exp}, we establish the exponential convergence
of the fidelity of CSS codes.
In Section~\ref{sec:BB84},
we apply Schumacher's argument to CSS codes to interpret a quantum code
as a QKD protocol,
and describe how this reduces to the BB84 protocol.
Section~\ref{ss:channel_para} 
reviews the method for channel parameter estimation in the BB84 protocol.
In Section~\ref{sec:security}, the security proof is completed.
Sections~\ref{ss:discussions} and \ref{sec:conc} contain discussions and the conclusion, respectively.
Proofs of subsidiary results are given in \ref{app:subsidiary}. 
In \ref{app:newdec}, an even better achievable rate, 
$1-\hmo(\erx)-\hmo(\erz)$, in the BB84 protocol is given.
A proof of security of a simple BB84-type protocol
for joint attacks is given in \ref{app:joint}.
The case of general joint attacks is treated 
in \ref{app:joint}.
A nomenclature can be found in \ref{app:nom}. 

\section{Calderbank-Shor-Steane Codes \label{sec:CSS}}

The complex linear space of operators on a Hilbert space $\Hch$ is
denoted by $\Bop(\Hch)$.
A quantum code usually means
a pair $(\cQ,\cR)$ consisting of a subspace $\cQ$
of $\Hch^{\tnsr n}$ and a trace-preserving completely positive
(TPCP) linear map $\cR$ on
$\Bop(\Hch^{\tnsr n})$, called a recovery operator; the subspace $\cQ$ alone
is also called a (quantum) code. 
Symplectic codes have more structure: They are
simultaneous eigenspaces of commuting operators on
$\Hch^{\tnsr n}$.
Once a set of commuting operators is specified,
we have a collection of eigenspaces of them.
A symplectic code refers to either such an eigenspace or a collection 
of eigenspaces, 
each possibly accompanied by a suitable recovery operator.
Hereafter, we assume $\Hch$ is a Hilbert space with an orthonormal
basis $\{ \ket{i} \}_{i=0}^{i=\dmn-1}$, and
$\dmn$ is a prime. 
Throughout, $\myF$ denotes $\SINT/\dmn\SINT$,
a finite field. 
We use the dot product defined by
\begin{equation}\label{eq:dotpr}
\dpr{(x_1,\dots,x_n)}{(y_1,\dots,y_n)}=\sum_{i=1}^{n} x_iy_i
\end{equation}
where the arithmetic is performed in $\myF$
(i.e., modulo $\dmn$ ),
and let $C^{\perp}$ denote $\{ y \in\myFpower{n} \mid \forall x\in C, \ \dpr{x}{y}=0 \}$
for a 
subset $\Ccl$ of $\myFpower{n}$.

In constructing symplectic codes, 
the following basis of $\Bop(\Hch^{\tnsr n})$ is used.
Let unitary operators $\Xbe, \Zbe$ on $\Hch$ 
be defined by
\begin{equation}\label{eq:error_basis}
\Xbe \ketbe{j}  = \ketbe{j-1}, \,\,\,
\Zbe \ketbe{j} = \phasebe^ j \ketbe{j}, \quad \,\,\, j\in\myF
\end{equation}
with $\phasebe$ being a primitive $\dmn$-th root of unity (e.g., $e^{\imu 2\pi/\dmn}$).
For $u=(u_1,\dots,u_n)\in\myFpower{n}$, let $\wbX{u}$ and $\wbZ{u}$ denote
$X^{u_1}\tnsr \cdots \tnsr X^{u_n}$
and $Z^{u_1}\tnsr \cdots \tnsr Z^{u_n}$, 
respectively.
The operators $\wbX{u}\wbZ{w}$, $u,w\in \myFpower{n}$,
form a basis of 
$\Bop(\Hch^{\tnsr n})$, which we call the Weyl (unitary)
basis~\cite{weyl28}.
Observe the commutation relation
\begin{equation}\label{eq:WCR}
(X^{u}Z^{w})(X^{u'}Z^{w'})=\omega^{\dpr{u}{w'}-\dpr{w}{u'}}
(X^{u'}Z^{w'})(X^{u}Z^{w}), \quad u,w,u',w'\in\myFpower{n},
\end{equation}
which follows from $XZ=\omega ZX$.
It is sometimes useful to rearrange the components of $(u,w)$ 
appearing in the operators $\wbX{u}\wbZ{w}$ in the Weyl basis as follows:
For $u=(u_1,\dots,u_n)$ and $w=(w_1,\dots,w_n) \in\myFpower{n}$,
we denote the rearranged one $\big((u_1,w_1),\dots,(u_n,w_n)\big) \in \cX^n$, 
where $\cX=\myF\times\myF$, 
by $[u,w]$.
We occasionally use another symbol $\Ebe$ for the Weyl basis:
$\Ebe_{[u,w]}=\wbX{u}\wbZ{w}$ and
$\Ebe_{\Ksp}= \{ \Ebe_x \mid x\in \Ksp \}$ for $\Ksp\in\cX^n$.

A CSS code is specified by a pair of classical linear 
codes (i.e., subspaces of $\myFpower{n}$) such that
one contains the other.
The quantum codes to be proved to have the desired performance in the sequel
are CSS codes of a special type, for which the pair is 
a classical code $\Ccl$ and its dual 
$\Ccl^{\perp}$ with the property
\begin{equation*}
\Csm \subset \Cbg.
\end{equation*}
This condition is equivalent to $\forall x,y\in \Ccl, \, \dpr{x}{y}=0$,
and a code $\Ccl$ satisfying it is said to be
{\em self-orthogonal}\/ (with respect to the dot product).

Coset structures are exploited in construction of CSS codes.
We fix some transversal (set of coset representatives
in which each coset has exactly one representative) of
the quotient group $\myFpower{n}/\Cbg$.
Identifying $\myFpower{n}/\Cbg$ and $\Cbg/\Csm$
with their fixed transversals, respectively,
we sometimes write, say, $x\in\myFpower{n}/\Cbg$ and $v\in\Cbg/\Csm$
for coset representatives $x$ and $v$.

Put $\kcl=\dim \Csm$, 
and assume $g_1,\dots, g_{\kcl}$ form a basis of $\Csm$.
The operators
\begin{equation}\label{eq:stab4CSS}
 \wbZ{g_1}, \dots, \wbZ{g_{\kcl}},\,\wbX{g_1}, \dots, \wbX{g_{\kcl}},
\end{equation}
commute with each other by (\ref{eq:WCR}) and $C \subset C^{\perp}$, so that
we have a collection of simultaneous eigenspaces of these operators,
which is called a CSS code.
Specifically, put
\begin{equation}\label{eq:encoded}
\ket{\phi_{xzv}} = \frac{1}{\sqrt{\crd{\Csm}}} \sum_{w\in \Csm}
\omega^{\dpr{z}{w}} \ket{w+v+x}
\end{equation}
for coset representatives
$x,z\in\myFpower{n}/\Cbg$ and $v\in\Cbg/\Csm$.
Then, we have
\begin{equation}\label{eq:eigen4stab}
\wbZ{g_j} \ket{\phi_{xzv}}= \omega^{\dpr{x}{g_j}} \ket{\phi_{xzv}}
\quad\mbox{and}\quad
\wbX{g_j} \ket{\phi_{xzv}}= \omega^{\dpr{z}{g_j}} \ket{\phi_{xzv}}, 
\quad j=1,\dots,\kcl .
\end{equation}
It is easy to check that
$\ket{\phi_{xzv}}$, $x,z\in\myFpower{n}/\Cbg,v\in\Cbg/\Csm$, form
an orthonormal basis of $\Hgen$.
In words, we have $\dmn^{n-2\kappa}$-dimensional subspaces $\cQ_{xz}$
such that $\bigoplus_{x,z} \cQ_{xz}=\Hch^{\tnsr n}$ and
$\cQ_{xz}$ is spanned by orthonormal vectors $\ket{\phi_{xzv}}$, 
$v\in \Cbg/\Csm$, for each pair $(x, z)\in(\myFpower{n}/\Csm^{\perp})^2$.
The subspaces $\cQ_{xz}$, $(x, z)\in(\myFpower{n}/\Csm^{\perp})^2$,
are the simultaneous eigenspaces of the operators in (\ref{eq:stab4CSS}),
and form a CSS code.

We will consistently use $\kcl$ and $k$ to denote 
$\kcl=\dim_{\myF} \Ccl$ and
\begin{equation}
k=n-2\kcl= \log_{\dmn} \dim_{\bC} \cQ_{xz}. \label{eq:k}
\end{equation}
Decoding or recovery operation for this type of CSS quantum codes is simple.
If we choose a transversal $\crJ$ of $\myFpower{n}/\Ccl^{\perp}$,
we can construct a recovery operator $\cR$ for $\cQ_{xz}$
so that the code $(\cQ_{xz},\cR)$ is $\Ebe_{\Kof{\crJ}}$-correcting
in the sense of \cite{KnillLaflamme97}, where 
\begin{equation} \label{eq:K}
\Kof{\crJ} = \{ [x,z] \mid x\in\crJ \mbox{ and } z\in\crJ \}.
\end{equation}
This directly follows from the general theory of symplectic
codes~\cite{crss97,crss98,gottesman96,hamada02c} on noticing that
the operators in the Weyl basis
that commute with all of those in (\ref{eq:stab4CSS}) are
$\wbX{u}\wbZ{w}$, $u\in\Ccl^{\perp}, w\in\Ccl^{\perp}$,
due to (\ref{eq:WCR}).
The $\Ebe_{\Kof{\crJ}}$-correcting CSS code specified by
$C$ and $\crJ$ as above will be denoted by $\css{C}{\crJ}$.

\section{Exponential Convergence of Fidelity of Codes to Unity \label{sec:exp}}

First, we treat the simple problem of establishing an attainable fidelity
of CSS codes.
We write
$P^{n}\big((x_1,\dots , x_n)\big)$ for $P(x_1) \cdots P(x_n)$
and $P^{n}(\Ksp)$ for 
$\sum_{ x\in \Ksp }P^{n}(x)$, where $P$ is a probability
distribution on $\cX$ and $\Ksp\subset\cX^n$.
More generally, $PQ$ denotes the usual product
of two probability distributions $P$ and $Q$,
which is specified by $[PQ](\vari,\varj)=P(\vari)Q(\varj)$.
For a probability distribution $Q$ on $\cY\times\cY$, we denote
the two marginal distributions by $\bar{Q}$ and $\dbar{Q}$:
\[
\bar{Q}(\vari)=\sum_{\varj\in\cY} Q(\vari,\varj),
\quad 
\dbar{Q}(\vari)=\sum_{\varj\in\cY} Q(\varj,\vari),
 \quad \ \vari \in \cY.
\]

\subsection{The case where $\dmn \ge 3$ \label{ss:exp3}}

The fidelity of the $\Ebe_{\Kof{\crJ}}$-correcting quantum code 
$\css{C}{\crJ}$ is 
not smaller than $P^n(\Kof{\crJ})$ when it is used on the quantum channel
that maps $\rho\in\Bop(\Hch^{\tnsr n})$ to 
$\sum_{x\in\cX^n} P^n(x) \Ebe_x \rho \Ebe_x^{\dagger}$.
This is true whether entanglement fidelity~\cite{schumacher96} 
or minimum fidelity~\cite{KnillLaflamme97} is employed.
This bound applies to general channels as well
(Section~\ref{ss:channel_para}).
Then, noticing
\begin{equation}\label{eq:PeXZ}
P^n(\Kof{\crJ}\cmple) \le \bar{P}^n(\crJ\cmple) + \dbar{P}^n(\crJ\cmple),
\end{equation}
where $\Ksp\cmple$ denotes the complement of $\Ksp$,
which holds by the definition (\ref{eq:K}) of $\Kof{\crJ}$, we will prove the following
theorem.

\begin{theorem} \label{th:main}
Assume $\dmn \ge 3$.
Let a number $0 \le R \le 1$ be given.
There exists a sequence of pairs 
$\{ (C\sqi{n}, \crJ\sqi{n}) \}$, each consisting
of a self-orthogonal code
$C\sqi{n} \subset \myFpower{n}$ with
$n-2 \dim_{\myF} C\sqi{n} \ge nR$ and a set $\crJ\sqi{n}$ 
of coset representatives 
of $\myFpower{n}/C\sqi{n}^{\perp}$, such that
for any probability distribution $P$ on $\cX=\myF\times\myF$, 
\[
 P^n(\Kof{\crJ\sqi{n}}\cmple) \le 
 \bar{P}^n(\crJ\sqi{n}\cmple) +  \dbar{P}^n(\crJ\sqi{n}\cmple) \le
\dmn^{-n E(R,\bar{P},\dbar{P}) + o(n) } 
\] 
where 
\[
E(R,\bar{P},\dbar{P})= \min\{ \Estar(R,\bar{P}), \Estar(R,\dbar{P}) \},
\]
\[
\Estar(R,\vp)= \min_{Q} [ D(Q||\vp) + 2^{-1} |1-2 H(Q)- R |^+ ],
\]
$|t|^+ =\max\{t,0\}$, $H$ and $D$ denote the entropy and the Kullback-Leibler
information with logarithms of base $\dmn$, respectively,
and the minimization with respect to $Q$ is taken 
over all probability distributions on $\myF$. 
\end{theorem}

{\em Remark 1.}\/
The function $E(R,\bar{P},\dbar{P})$ is strictly positive for $R<1-
2\max\bigl\{H(\bar{P}),H(\dbar{P})\bigr\}$.
The 
code $\css{C\sqi{n}}{\crJ\sqi{n}}$ has rate $1-2 \dim_{\myF}
C\sqi{n}/n \ge R$.
The code $C^{\perp}_n$, as a classical channel code of rate not less than
$R'=(R+1)/2$,
attains the error exponent $\Estar(2R'-1,\vp)$ 
known as the random coding error exponent~\cite{CsiszarKoerner} 
of the memoryless additive channel
that changes an input $a\in\myF$ into $a-b$ with probability $p(b)$.

{\em Remark 2.}\/
Whereas 
$P^n(\Kof{\crJ\sqi{n}})$
is a measure of the performance of quantum code $\css{\Ccl_n}{\crJ_n}$,
the probability $\bar{P}^n(\crJ\sqi{n}\cmple)$
has its own meaning. It 
is an upper bound on 
the probability of decoding error for the key transmission,
which is proved in \ref{app:subsidiary}. 
In fact, the error probability is
$\bar{P}^n(\crJ\sqi{n}\prcmple)$ where $\crJ_n'=\crJ_n+\Ccl_n$,
not
$\bar{P}^n(\crJ\sqi{n}\cmple)$
because adding a word $e$ in $\Ccl_n$ to the key $v+\Ccl_n$
does not change it.

{\em Remark 3.}\/
That $\crJ_n'$ is the effective correctable errors
in QKD (Remark~2) may be interpreted as a manifestation
of an inherent property, which is sometimes called `degeneracy', of CSS codes 
(more generally, of symplectic codes):
Put $\crJ'=\crJ+\Ccl$; Then, a CSS code $\css{\Ccl}{\crJ}$,
as a quantum code, 
can correct the `errors' 
$\Ebe_{y}$, $y\in\Kof{\crJ'}$%
~\cite{crss97,crss98} (or e.g., \cite{hamada02c,hamada03f}). 

{\em Remark 4.}\/
The function $o(n)$ is explicitly given as $3(\dmn-1)\log_{\dmn}(n+1)+\log_{\dmn} 2 + \dmn$ by (\ref{eq:prooflast}) below.


We prove the theorem by a random coding argument,
which is 
analogous to that in \cite{hamada01e}, 
where the idea of universal decoding,
i.e., minimum entropy (maximum mutual information)
decoding of Goppa (e.g., \cite{CsiszarKoerner,CsiszarKoerner81a}) was
already used.
For the present purposes, we want the codes $\Ccl_n$ also to be robust
or 
universal in the sense that their structures do not
depend on the distribution $P$, which characterizes the channel. 
To show this, we begin with the next lemma, 
which is a variant of Calderbank and
Shor's~\cite[Section~V]{CalderbankShor96}
and says that 
the ensemble of all self-orthogonal codes is `balanced'. 

\begin{lemma}\label{lem:C_uniform}
Assume $\dmn \ge 3$, and
let
\[
\Aso = \Aso^{(n,\kappa)} = \{ \Ccl \subset \myFpower{n} \mid \mbox{$\Ccl$ {\rm linear}}, \ \Ccl \subset
\Ccl^{\perp},\ \dim \Ccl = \kcl \}
\]
and
\[
\Acn{x} = \big\{ \Ccl \in \Aso \mid x \in \Ccl^{\perp} \big\}.
\]
Then, for any $\varaaa\in\myF$, there exists a constant $\Nwd{\varaaa}$ such that 
$\crd{\Acn{x}}=\Nwd{\varaaa}$ for any non-zero word $x\in\myFpower{n}$ with $\dpr{x}{x}=\varaaa$.
\enlem

{\em Remark}.\/
The proof below is the same as that of Lemma~6 in \cite{hamada01g}
except that the dot product is used here
in place of the standard symplectic form. This is possible
because $\myFpower{n}$ equipped with the dot product is 
an orthogonal space if $\dmn$ is a prime other than $2$. 
The case of $\dmn=2$ is exceptional, and will be treated later.
Lemma~\ref{lem:C_uniform} and the corollary below are true if
the dot product is replaced by any orthogonal,
symplectic or unitary form more generally.

{\em Proof.}\/ 
To prove $\crd{\Acn{x}}=\crd{\Acn{y}}$ for non-zero vectors $x$ and $y$
with $\dpr{x}{x}=\dpr{y}{y}=\varaaa$,
it is enough to show the existence of an isometry $\alpha$
(an invertible linear map $\alpha$ that preserves the `product',
i.e., that satisfies $\dpr{\alpha(x)}{\alpha(y)}=\dpr{x}{y}$
for all $x$ and $y$)
on $\myFpower{n}$ with $y=\alpha(x)$,
but this directly follows 
from the well-known Witt lemma~\cite{artin,serre,aschbacher,grove}, which
states that 
any isometry that is defined on a subspace of an orthogonal space $V$
can be extended to an isometry on the whole space $V$.
\enproof

\begin{corollary}\label{coro:C_uniform}
For $x\in\myFpower{n}$, $\dmn \ge 3$,
\[
\frac{\crd{\Acn{x}}}{\crd{\Aso}} \le \left\{ \begin{array}{cc}
\dmn^{-\kcl+\dmn-1} & \mbox{\rm if \ $x\ne \zrv$}\\
1 & \mbox{\rm if \ $x=\zrv$.} 
\end{array} \right.
\]
\end{corollary}

{\em Proof.}\/ 
The case of $x=\zrv$ is trivial.
Let $S_{\varaaa}=\crd{ \{ x \in\myFpower{n} \mid \dpr{x}{x}=\varaaa, x \ne \zrv \} }$
for $\varaaa\in\myF$.
Counting the pairs $(x,\Ccl)$ such that $x\in\Ccl^{\perp}$, 
$\dpr{x}{x}=\varaaa$, $x\ne \zrv$ and $C\in\Aso$
in two ways, we have $S_{\varaaa}\Nwd{\varaaa}\le\crd{\Aso}(\dmn^{n-\kcl}-1)$.
But $S_{\varaaa} \ge \dmn^{n-\dmn+1}-1$ (since $x \in S_u$ can take 
arbitrary values in the first $n-\dmn+1$ positions except $(0,0, \dots, 0)$), 
and hence we have $(\dmn^{n-\dmn+1}-1) \Nwd{\varaaa} \le \crd{\Aso}(\dmn^{n-\kcl}-1)$,
from which the desired estimate follows.
\enproof

In the proof of Theorem~\ref{th:main},
we will use the method of types, a standard tool in information theory.
Here we collect the needed notions and basic inequalities regarding 
the method of types.
With a finite set $\cY$ fixed, 
the set of all probability distributions on $\cY$ is denoted by $\cP(\cY)$.
The {\em type}\/ of a sequence $y=(y_1,\dots,y_n)\in\cY^n$,
denoted by $\sP_{y}$,
represents the relative frequencies of appearances of symbols $\vari\in\cY$ in $y$:
\begin{equation}\label{eq:type}
\sP_{y}(\vari)=\frac{\crd{\{ i \mid 1\le i \le n, y_i = \vari \}}}{n}, \quad \vari \in \cY.
\end{equation}
The set of all possible types of sequences in $\cY^n$
is denoted by $\cP_n(\cY)$, 
and for $Q\in\cP_n(\cY)$, the set of sequences of type $Q$
and length $n$ is denoted by $\cT_{Q}^n$ or $\cT_{Q}^n(\cY)$.
In what follows, we use 
\begin{equation}\label{eq:types}
\crd{\cP_n(\cY)}\le (n+1)^{\dmn-1},\quad \mbox{and}\quad\forall Q\in \cP_n(\cY),\
|\cT_{Q}^n| \le \dmn^{nH(Q)},
\end{equation}
where $\dmn=\crd{\cY}$.
Note that if $x\in\cY^{n}$ has type $Q$, then
$\vp^{n}(x)=\prod_{\vari\in\cY} \vp(\vari)^{nQ(\vari)} = 
{\dmn}^{ -n [H(Q)+D(Q||\vp)] }$ for any $p\in\cP(\cY)$,
so that the probability that words of a fixed type $Q\in\cP_n(\cY)$
occur has the bound
\begin{equation}\label{eq:prob_type}
\sum_{y\in\cY^n: \sP_y=Q} \vp^{n}(\vxy) \le \dmn^{-nD(Q||\vp)}.
\end{equation}

Now we are ready to prove the existence of a `balanced' code,
which will turn out to be universal. 
Given a set $C'\subset\myFpower{n}$, put 
\begin{eqnarray*}
\tsptwo{Q}{C'}&=& \crd{\{ x\in C' \mid \sP_x = Q \}} \\
&=&\sum_{x\in\myFpower{n}} \chf[ x\in C' \mbox{ and } \sP_x = Q ],
\quad Q \in\cP_n(\myF),
\end{eqnarray*}
where $\chf[T]$ equals $1$ if the statement $T$ is true and equals $0$
otherwise, and put
\[
\tspbarone{Q} =\frac{1}{\crd{\Aso}}\sum_{C\in\Aso} \tsptwo{Q}{C^{\perp}}.
\]
Then,
we obtain the next lemma following the method in \cite{ABargKnillL00II} (cf.\ \cite{barg02}).
\begin{lemma}\label{lem:goodcode}
For any $n \ge 2$ and $\kappa \le n/2$,
there exists a code $C$ in $\Aso=\Aso^{(n,\kappa)}$ such that
\[
\forall Q\in\cP_n(\myF), \quad \tsptwo{Q}{C^{\perp}} \le \crd{\cP_n(\myF)} \tspbarone{Q}.
\]
\end{lemma}
{\em Remark.}\/ The list of numbers
$\big(\tsptwo{Q}{C'}\big)_{Q\in\cP_n(\myF)}$, type spectrum, so to speak, is 
a natural generalization of 
the weight spectrum (distribution) in coding theory.
In fact, they are the same when $\dmn=2$.

\proof
Regarding $C$ as a random variable uniformly distributed over $\Aso$
and using Markov's inequality (e.g., \cite{CoverThomas}), 
which states that $\Prob\{ \sX \ge a \mu \} \le 1/a$
for a positive constant $a$, 
and a random variable $\sX$ that takes non-negative values and has
a positive mean $\mu$, 
we have
\begin{eqnarray*}
\lefteqn{\Prob \{ \exists Q \in \cP_n(\myF),\, \tsptwo{Q}{C^{\perp}} \ge \crd{\cP_n(\myF)}^{1+\vep}  \tspbarone{Q} \mbox{ and } \tspbarone{Q} > 0 \} }\\
\le \sum_{Q\in\cP_n(\myF):\, \tspbarone{Q} > 0 } \Prob \{ \tsptwo{Q}{C^{\perp}} \ge \crd{\cP_n(\myF)}^{1+\vep}  \tspbarone{Q} \} \le 1/\crd{\cP_n(\myF)}^{\vep}
\end{eqnarray*}
for any $\vep>0$.
Hence, $\Prob \{ \forall Q \in \cP_n(\myF),\, \tsptwo{Q}{C^{\perp}} < \crd{\cP_n(\myF)}^{1+\vep}  \tspbarone{Q} \mbox{ or } \tspbarone{Q}=0 \} \ge 1-1/\crd{\cP_n(\myF)}^{\vep}>0$. Since $\vep>0$ is arbitrary, this implies
the lemma.
\enproof

\begin{corollary}\label{coro:goodcode}
There exists a code $C$ in $\Aso=\Aso^{(n,\kappa)}$ such that
for any $Q\in\cP_n(\myF)$, $Q\ne\sP_{0^n}$,
\[
\frac{\tsptwo{Q}{C^{\perp}}}{\crd{\cT_{Q}^n}}
\le \crd{\cP_n(\myF)}\wkb \dmn^{-\kappa+\dmn-1}. 
\]
\end{corollary}
\proof
We have
\begin{eqnarray}
\tspbarone{Q} &=&\frac{1}{\crd{\Aso}}\sum_{C\in\Aso} 
\sum_{x\in\myFpower{n}} \chf[ x\in C^{\perp} \mbox{ and } \sP_x = Q ] \nonumber\\
&=&\sum_{x\in\myFpower{n}:\,\sP_x=Q} \frac{1}{\crd{\Aso}}\sum_{C\in\Aso}
\chf[ x\in C^{\perp} ]\nonumber\\
&\le & \sum_{x\in\myFpower{n}:\,\sP_x=Q} \dmn^{-\kappa+\dmn-1}\nonumber\\
&= & \crd{\cT_{Q}^n}  \dmn^{-\kappa+\dmn-1}, \quad Q \ne \sP_{0^n},\label{eq:type4A}
\end{eqnarray}
where the inequality is due to Corollary~\ref{coro:C_uniform}, and hence the desired estimate.
\enproof

Corollary~\ref{coro:goodcode} says that
there exists a code $C\in\Aso$ such that
$\big(\tsptwo{Q}{C^{\perp}}\big)_{Q\in\cP_n(\myF)}$ is almost proportional to
$\big(\tsptwo{Q}{\myFpower{n}}\big)_{Q\in\cP_n(\myF)}
=(\crd{\cT_{Q}^n} )_{Q\in\cP_n(\myF)}$.
[Clearly, the code $C^{\perp}$ in this corollary satisfies the Gilbert-Varshamov bound asymptotically.]
We will see that the code in Lemma~\ref{lem:goodcode} or Corollary~\ref{coro:goodcode} has the universality mentioned above. 

The decoding should also possess such universality. 
Note that for CSS codes, in theory, 
the design of a decoder is accomplished by choosing a transversal of
$\myFpower{n}/\Ccl^{\perp}$.
Based on the idea of minimum entropy decoding, 
{\em from each of the $\dmn^{\kcl}$ cosets of $\Ccl^{\perp}$ in $\myFpower{n}$,
we choose a vector that minimizes $H(\sP_{x})$ in the coset.}
To break ties, we use an arbitrarily fixed order, say a lexicographic order
in $\myFpower{n}$.

{\em Proof of Theorem~\ref{th:main}.}\/
In the proof, $\cP_n(\myF)$ is abbreviated as $\cpf$.
Fix a code $\Cgood$ of the property in Corollary~\ref{coro:goodcode}
and a transversal $\Jgood$ chosen as above. We will show $\Cgood$ is the
desired code.
Let $\cS_n$ be the group composed of all
permutations on $\{ 1,\dots, n\}$ and assume $\prm\in\cS_n$, when applied to $\Cgood$ or $\Jgood$, permutes all words in $\Cgood$ or $\Jgood$ as
$\prm([x_1,\dots,x_n])=[x_{\prm(1)},\dots,x_{\prm(n)}]$.
Clearly, $p^n(\Jof{\Ccl})=p^n(\Jgood)$ for any $\prm\in\Ensperm$
and any probability distribution $p$ on $\myF$.
For a technical reason, we will evaluate the ensemble average of 
$p^n(\Jof{\Ccl})$ over $\Ensperm$, 
which equals $p^n(\Jgood)$, the original quantity in question.
Specifically,
put
\begin{equation}\label{eq:avP}
\Bp{\vp} = \frac{1}{\crd{\Ensperm}} \sum_{\prm\in\Ensperm} \vp^n(\Jof{\Ccl}\cmple)
\end{equation}
for $\vp=\bar{P},\dbar{P}$.
We will show, for some polynomial $f(n)$, that $\Bp{\vp}$ is bounded 
above by $f(n) \dmn ^{ - n \Estar(R,\vp) }$,
which implies
$\Bp{\bar{P}}+\Bp{\dbar{P}} \le 2f(n) \dmn ^{ - n \min \{
\Estar(R,\bar{P}),\Estar(R,\dbar{P}) \} }$.
This, together with (\ref{eq:PeXZ}), establishes the theorem.

It was shown that
an exponential fidelity bound holds for a `balanced' ensemble of additive codes~\cite{hamada01e,hamada01g}.
To take the same approach as in \cite{hamada01e,hamada01g}, 
we show that the ensemble
$\prm(\Cgood)^{\perp}$, $\prm\in\Ensperm$, is almost `balanced'.
Imagine we list up all words in $\prm(\Cgood)^{\perp}$ for all $\prm\in\Ensperm$.
Clearly, for any $Q\in\cpf$, 
there exists a constant, say $\cnt{Q}$, such that
$\crd{\{\prm\in\Ensperm \mid x\in \prm(\Cgood)^{\perp} \}}=\cnt{Q}$
for any word $x$ with $\sP_x=Q$. 
Then, counting the number of words of a fixed type $Q$ in the list in
two ways,
we have $\crd{\cT_{Q}^n} \cnt{Q} = \crd{\Ensperm} \tsptwo{Q}{\Cgood^{\perp}}$.
Hence, for any type $Q \ne \sP_{0^n}$,
\begin{equation}
\frac{\cnt{Q}}{\crd{\Ensperm}}=\frac{\tsptwo{Q}{\Cgood^{\perp}}}{\crd{\cT_{Q}^n}}
\le \crd{\cpf}\wkb \dmn^{-\kappa+\dmn-1}, \label{eq:uniQ}
\end{equation}
where we have used Corollary~\ref{coro:goodcode}.
We have proved the next lemma.
\begin{lemma}\label{lem:balanced_perm}
Put
\[
\Acn{y}(\Cgood) = \big\{ \prm \in \Ensperm \mid y \in \prm(\Cgood)^{\perp} \big\}.
\]
For $y\in\myFpower{n}$, $y\ne 0^n$, we have
\[
\frac{\crd{\Acn{y}(\Cgood)}}{\crd{\Ensperm}} \le \crd{\cP_n}\wkb \dmn^{-\kcl+\dmn-1}.
\]
\end{lemma}

From (\ref{eq:avP}), we have
\begin{eqnarray}
\Bp{\vp} & = & \frac{1}{\crd{\Ensperm}} \sum_{\prm\in\Ensperm} \sum_{\vxy \notin \Jof{\Ccl}}  \vp^n(\vxy) \nonumber\\
     & = & \sum_{x \in \myFpower{n}} \vp^n(x) \frac{ \crd{\{ \prm \in
     \Ensperm \mid
     x \notin \Jof{\Ccl} \}} }{\crd{\Ensperm}}.\label{eq:pr0}
\end{eqnarray}
Since $x \notin \Jof{\Ccl}$ occurs only if there
 exists
a word $u\in\myFpower{n}$ such that $H(\sP_u) \le H(\sP_x)$ and $u-x\in
 \prm(\Cgood)^{\perp} \setminus \{ \zrv \} $
from the design of $\Jgood$
specified above (minimum entropy decoding),
it follows
\begin{eqnarray}
\lefteqn{ \!\!\! \crd{ \{ \prm \in \Ensperm \mid x \notin \Jof{\Ccl} \}}/\crd{\Ensperm} }\nonumber\\
 &\le &\sum_{u\in \myFpower{n} :\, H(\sP_u) \le  H(\sP_x),\ u \ne x}
 \crd{\Acn{u-x}(\Cgood)}/\crd{\Ensperm}\nonumber\\
  &\le & \sum_{u\in \myFpower{n} :\, H(\sP_u) \le  H(\sP_x)} 
\crd{\cpf}\wkb \dmn^{-(\kcl-\dmn+1)}, \nonumber\\
 & = & \sum_{Q'\in \cpf:\, H(Q') \le  H(\sP_x)}
\crd{\cpf}\wkb \crd{\cT_{Q'}^n}\dmn^{-(\kcl-\dmn+1)} \nonumber\\
 & \le & \sum_{Q'\in \cpf:\, H(Q') \le  H(\sP_x)}
\crd{\cpf}\wkb \dmn^{nH(Q')-(\kcl-\dmn+1)} 
\label{eq:pr2} 
\end{eqnarray}
where we have used 
Lemma~\ref{lem:balanced_perm} for the second
inequality, and (\ref{eq:types}) for the last inequality.
Then, recalling (\ref{eq:k}) and (\ref{eq:prob_type}),
and choosing the smallest integer $k$ such that
$k \ge nR$ and $\kappa=(n-k)/2$ is an integer, which implies $nR \le k< nR+2$,
with repeated use of the inequality
$\min \{ \tvara+\tvarb, 1\} \le \min \{ \tvara, 1\} + \min \{ \tvarb, 1\}$ for $\tvara,\tvarb \ge 0$,
we can proceed from (\ref{eq:pr0}) as follows:
\begin{eqnarray*}
\Bp{\vp} &\le& \sum_{x\in\myFpower{n}} \vp^n(x) \min \Biggl\{
\sum_{ Q'\in \cpf:\, H(Q') \le
H(\sP_x) } \crd{\cpf}\wkb \dmn^{nH(Q')-(\kcl-\dmn+1) } 
,\ 1 \ \Biggr\}\\
 &\le &  \crd{\cpf}\wkb \sum_{Q\in\cpf} {\dmn}^{-nD(Q||\vp)+\dmn} 
\min \Biggl\{
\sum_{Q'\in \cpf:\, H(Q') \le
H(Q)} \dmn^{nH(Q')-\frac{n-k}{2}-1}  
, \ 1 \ \Biggr\}\\
 &\le &  \crd{\cpf}\wkb \sum_{Q\in\cpf} {\dmn}^{-nD(Q||\vp)+\dmn}
\sum_{Q'\in \cpf:\, H(Q') \le
H(Q)} \min \bigl\{ \dmn^{-n[1-R-2H(Q')]/2}  
,\ 1 \ \bigr\}\\
&\le & \crd{\cpf}^2 \sum_{Q\in\cpf} {\dmn}^{-nD(Q||\vp)+\dmn} 
\max_{Q'\in \cP(\myF):\, H(Q') \le
H(Q) } {\dmn}^{- n |1-R-2H(Q')|^+/2 }\\
&=&  \crd{\cpf}^2  \sum_{Q\in\cpf} {\dmn}^{-nD(Q||\vp)+\dmn} 
{\dmn}^{-  n 
|1-R-2H(Q) |^+/2 }\\
&\le& \dmn^{\dmn} \crd{\cpf}^3 \max_{Q}
{\dmn}^{ 
- n [D(Q||\vp)+
|1-R-2H(Q)|^+/2 ] } 
= \dmn^{\dmn} \crd{\cpf}^3 \dmn^{-n \Estar(R,p)}.
\label{eq:lastchain}
\end{eqnarray*}
Hence, we have
\begin{eqnarray}
\Bp{\bar{P}}+\Bp{\dbar{P}} & = & 
\frac{1}{\crd{\Ensperm}} \sum_{\prm\in\Ensperm} 
[\bar{P}^n(\Jof{\Ccl}\cmple) + \dbar{P}^n(\Jof{\Ccl}\cmple)] \nonumber\\
& \le & 2 \dmn^{\dmn} \crd{\cpf}^3
 \dmn ^{ - n \min \{
\Estar(R,\bar{P}),\Estar(R,\dbar{P}) \} }. \label{eq:prooflast}
\end{eqnarray}
Since $\crd{\cpf}\le (n+1)^{\dmn-1}$,
we obtain the desired bound.
\enproof

\subsection{The case where $\dmn =2$ \label{ss:exp2}}

Calderbank and Shor~\cite{CalderbankShor96} proved
the following lemma based on a result in coding theory.
\begin{lemma}\label{lem:C_uniform2}
Assume $\dmn=2$,
$n \ge 2$ is an even integer, and $0< \kcl \le n/2$ is an integer. 
Let 
\[
\Aso = \{ \Ccl \subset \myFpower{n} \mid \mbox{$\Ccl$ {\rm linear}}, \ \{ 1^n \} \subset \Ccl \subset
\Ccl^{\perp},\ \dim \Ccl = \kcl \},
\]
and
\[
\Acn{x} = \big\{ \Ccl \in \Aso \mid x \in \Ccl^{\perp} \big\}.
\]
Then, there exists a constant $\Nwd{0}$ satisfying
$\crd{\Acn{x}}=\Nwd{0}$
for any $x\in\myFpower{n}$ with $\dpr{x}{x}=0$,
$x\ne \zrv$ and $x \ne 1^n$.
\enlem

\begin{corollary}\label{coro:C_uniform2}
For $x\in\myFpowerarg{2}{n}$,
\[
\frac{\crd{\Acn{x}}}{\crd{\Aso}} \le \left\{ \begin{array}{cc}
\dmn^{-\kcl+\dmn-1} & \mbox{\rm if \ $x \ne \zrv$ and $x \ne 1^n$}\\
1 & \mbox{\rm if \ $x=\zrv$ or $x=1^n$.} 
\end{array} \right.
\]
\end{corollary}

{\em Remark.}\/ Trivially, $\crd{\Acn{x}}=0$ for all $x$ with $\dpr{x}{x}=1$
since $\dpr{x}{x}=\dpr{x}{1^n}$.
We can also prove this lemma noticing
a hidden structure of a symplectic space.
Namely,
letting $\sF_{\rm even}$ be the set of of all 
words $x$ with $\dpr{x}{x}=0$ in $\myFpowerarg{2}{n}$,
and noting that the additive quotient group $\sF_{\rm even}/\spn 1^n$,
where $\spn 1^n =\{ 0^n, 1^n \}$, 
is a symplectic space equipped with the natural form 
$\dpr{(x+\spn 1^n)}{(y+\spn 1^n)}=\dpr{x}{y}$,
we can argue as in the proof of Lemma~\ref{lem:C_uniform}.

In Theorem~\ref{th:main}, due to Remark~3 thereof, we could have used $\crJ'$
or a subset $\tlJ$ of $\crJ'$
in place of $\crJ$ for the purposes of evaluating the fidelity
(and the probability of disagreement between Alice's key and Bob's
due to Remark~2 to Theorem~\ref{th:main}).
Namely, we obtain Theorem~\ref{th:main} with `$\dmn \ge 3$' and
`$P^n(\Kof{\crJ\sqi{n}}\cmple) \le 
 \bar{P}^n(\crJ\sqi{n}\cmple) +  \dbar{P}^n(\crJ\sqi{n}\cmple)$'
replaced by `$\dmn=2$ and $n$ is even' and 
`$P^n(\Kof{\tlJ\sqi{n}}\cmple) \le 
 \bar{P}^n(\tlJ\sqi{n}\cmple) +  \dbar{P}^n(\tlJ\sqi{n}\cmple)$',
respectively, where $\tlJ\sqi{n}=\crJ\sqi{n}+1^n$,
and using Corollary~\ref{coro:C_uniform2} instead of 
Corollary~\ref{coro:C_uniform} in the above proof of
Theorem~\ref{th:main}. 
In fact, with $\crJ$ replaced by $\tlJ$,
the proof of Theorem~\ref{th:main} can read
verbatim except the first inequality in (\ref{eq:pr2}),
which should be replaced by
\[
\crd{ \{ \prm \in \Ensperm \mid x \notin \prm(\tlJ) \} }
\le \sum_{u\in \myFpower{n} :\, H(\sP_u) \le  H(\sP_x),\ u-x \ne 0^n,1^n}
 \crd{\Acn{u-x}(\Cgood)},
\]
and the other few words. Thus, the statement of Theorem~\ref{th:main} 
is true for $\dmn=2$ with $\crJ$ replaced by $\tlJ$ and with the restriction
of $n$ being even, where the code $C_n$ always contains $1^n$.
[For $\dmn=2$ and $n$ odd, a geometric argument based on isometries
as before shows that 
the rate $1-2\hmo\big((\erx+\erz)/2\big)$ is achievable
for $(\erx+\erz)/2 \le 1/2$, whereas
the restriction $(\erx+\erz)/2 \le 1/2$ is not needed for $n$ even.
In this case, we use isometries on $\myFpower{n}$ that fix $1^n$,
with respect to the dot product,
noticing that $\myFpower{n}=\sF_{\rm even}+\spn 1^n$
and $1^n$ is orthogonal to $\sF_{\rm even}$ in order to prove the existence of balanced codes; we use the minimum Hamming distance decoding in place of the minimum entropy decoding.]


\section{Bennett-Brassard 1984 Quantum Key Distribution Protocol \label{sec:BB84}}

In the proof of the security of the BB84 protocol,
Shor and Preskill used
the observation of Lo and Chau~\cite{LoChau99}, 
who upper-bounded the amount of information 
that the eavesdropper, Eve, could obtain on the key 
by the Holevo bound~\cite{holevo73}.
However, a similar observation using the Holevo bound
had already been made by Schumacher~\cite[Section~V-C]{schumacher96},
who directly related Eve's information with quantum channel codes.
In this section, 
we will apply Schumacher's argument
to CSS codes
to avoid a detour to entanglement distillation.

\subsection{Quantum Codes and Quantum Cryptography \label{ss:principle_sch}}

Suppose we send a $k$-digit key $\rvv+\Csm \in \Cbg/\Csm$ encoded
into $\ket{\phi_{\rvx\rvz\rvv}}\in\cQ_{\rvx\rvz}$, 
where we regard $\rvx,\rvz,\rvv$ as random variables,
and $(\rvx,\rvz,\rvv)$ are randomly chosen
according to the uniform distribution. 
Once Eve has done an eavesdropping, namely, a series of measurements,
Eve's measurement results form another random variable, say, $\rve$.
We use the standard symbol $I$ to denote the mutual information (\ref{app:nom}). 

According to \cite[Section~V-C]{schumacher96}, 
\begin{equation}\label{eq:Sch1}
I(\rvv;\rve|\rvx=x,\rvz=z) \le \Se{x}{z}
\end{equation}
where $\Se{x}{z}$ is the entropy exchange after the
system suffers a channel noise $\cN$, Eve's attack $\cE$, another
channel noise $\cN'$, and the recovery
operation $\cR=\cR_{xz}$ for $\cQ_{xz}$ at the receiver's end.
Let us denote by $F_{xz}$ the fidelity of the code $(\cQ_{xz},\cR)$
employing the entanglement fidelity $\Fen$~\cite{schumacher96}.
Specifically,
\[
F_{xz}=\Fen\big(\pi_{\cQ_{xz}}, \cR\cN'\cE\cN\big)
\]
where $\pi_{\cQ}$ denotes the normalized projection operator onto $\cQ$,
and $\cB\cA(\rho)=\cB\big(\cA(\rho)\big)$ for two CP maps $\cA$ and $\cB$,
etc.
Then, by the quantum Fano inequality~\cite[Section~VI]{schumacher96}, we have
\begin{equation}\label{eq:Sch2}
\Se{x}{z} \le h(F_{xz}) + (1-F_{xz}) 2 nR
\end{equation}
where $R=n^{-1}\log_{\dmn}\dim\cQ_{xz}$.
Combining (\ref{eq:Sch1}) and (\ref{eq:Sch2})
and taking the averages
of the end sides, we obtain
\begin{eqnarray}
I(\rvv;\rve|\rvx\rvz) 
&\le & \Expe h(F_{\rvx\rvz}) + (1-\Expe F_{\rvx\rvz})2 nR\nonumber\\
&\le & h(\Expe F_{\rvx\rvz}) + (1-\Expe F_{\rvx\rvz}) 2 nR,\label{eq:Sch3}
\end{eqnarray}
where $\Expe$ denotes the expectation operator with respect to $(\rvx,\rvz)$.
Hence, if $1-\Expe F_{\rvx\rvz}$ 
goes to zero faster than $1/n$, then $I(\rvv;\rve|\rvx\rvz) \to 0$
as $n\to\infty$. We have seen that the convergence is, in fact,
exponential for some good CSS codes,
viz., $1-\Expe F_{\rvx\rvz} \le \dmn^{-n E + o(n)}$ with some $E>0$.
This, together with (\ref{eq:Sch3}), implies 
\begin{equation}
I(\rvv;\rve|\rvx\rvz) \le 
2\dmn^{-nE+o(n)} [n(E  +  R) -o(n)], \label{eq:Sch5}
\end{equation}
where we used the upper bound $- 2 t \log t$ for $h(t)$, $0\le t \le1/2$,
which can easily be shown by differentiating $t\log t$
(or by Lemma~2.7 of \cite{CsiszarKoerner}).
Thus, we could safely send a key $v+\Csm$ 
provided we could send
the entangled state $\ket{\phi_{xzv}}$ in (\ref{eq:encoded})
and the noise level of 
the quantum channel including Eve's action were tolerable by the
quantum code.

\subsection{Reduction to the Bennett-Brassard 1984 Protocol}

To reduce the above protocol to a more practical one, namely the BB84 protocol,
we use 
Shor and Preskill's observation that 
the probabilistic mixture of $\ket{\phi_{xzv}}$ 
with $x,v$ fixed and $z$ chosen uniformly randomly 
over $\myFpower{n}/C^{\perp}$
is given as
\begin{equation}\label{eq:SPmixed}
\frac{1}{\crd{C}}\sum_{z}\ket{\phi_{xzv}}\bra{\phi_{xzv}} = \frac{1}{\crd{C}}\sum_{w\in C}\ket{w+v+x}\bra{w+v+x},
\end{equation}
which can be prepared as the mixture of states 
$\ket{w+v+x}$ with no entanglement.
%
Then, it is seen that sending the key $v$ encoded into the state 
in (\ref{eq:SPmixed}) with $x$ chosen randomly is exactly what is done 
in the following protocol of Bennett and Brassard, 
which is essentially the same as
that in \cite{ShorPreskill00} except
that a CSS code of a higher rate is chosen in Step (vii). 

In the protocol, introduced are three more sequences of independent and
identically distributed binary random variables
$\rva,\rvb,\rvc$, where $\rva=(\rva_1,\dots,\rva_{\varn})$ and so on.
The probability of occurrence of 1 for the bits of $\rva,\rvb,\rvc$  
will be denoted  by $\pa$, $\pb$, $\pc$, respectively, where
$\pa,\pb,\pc\in (0,1)$.
We put
\begin{equation}\label{eq:code2whole}
\vra=\frac{\pa\pb}{\pa\pb+(1-\pa)(1-\pb)},
\end{equation}
which is the expected ratio of the number of $i$'s
with $a_i=b_i=1$ to that of $i$'s with $a_i=b_i$.
%
In what follows, the $Z$-basis denotes the collection $\ket{j}$, $j\in\myF$,
the $Z$-basis measurement denotes the simple (projective)
measurement $\{ \ket{j}\bra{j} \}_j$.
We also say `measure $Z$' in place of `perform the $Z$-basis measurement'.
The $X$-basis,  $X$-basis measurement, and `measure $X$'
are to be similarly understood with the $\dmn$ orthogonal eigenstates of $X$.
Specifically, the $X$-basis consists of
\[
\ket{j}' = \sum_{\vara\in\myF} \omega^{j\vara} \ket{\vara}, \quad j\in\myF.
\]

\begin{center}
{\bf BB84 protocol}
\end{center}
\begin{enumerate}
\item   The sender, Alice, and the receiver, Bob, do Steps (ii)--(iv) 
for each $i=1,\dots, \varn$.

\item   Alice chooses a random bit $\rva_i$.
        She prepares her system in one state that is chosen uniformly randomly from the $Z$-basis
	if $\rva_i$ is $0$, or
        in one from the $X$-basis if $\rva_i$ is $1$.

\item Alice sends the prepared state to Bob.

\item Bob chooses another random bit $\rvb_i$,
      and receives the state, 
      performs the $Z$-basis measurement
      if $\rvb_i$ is $0$, or 
      $X$-basis measurement if $\rvb_i$ is $1$.
	
\item Alice and Bob announce $\rva=(\rva_1,\dots,\rva_{\varn})$ and $\rvb=(\rvb_1,\dots,\rvb_{\varn})$, respectively.

\item Alice and Bob discards any results where $\rva_i \ne \rvb_i$.
      Alice 
      draws another string of random bits 
      $\rvc=(\rvc_1,\dots,\rvc_{\varn})$, and sends it
      to Bob through a public channel.
      They decide that those $\dmn$-ary digits 
      with the accompanying $\rvc_i$ being 0
      will be the code digits,
      i.e., will be used for key transmission with a CSS code.
      In the case where $\dmn=2$,
      it is assumed that the number of the code digits is even (if not, they divert
      one digit chosen in an arbitrary 
      manner to estimation of the noise level in the following step).
\item Alice and Bob announce the values of their non-code digits
which are accompanied by $\rvc_i=1$,
and from these and $\rva_i$ ($=\rvb_i$), estimate the noise level, and decide
on a secure transmission rate, and a CSS code, i.e., a pair $(C,\crJ)$, to be used
(the exact meaning will be clear in Section~\ref{sec:security}).

\item Alice announces the coset $y+\Cbg$, where $y$ ($=w+v+x$) is the string
     consisting of the remaining code digits. In other words,
     she announces 
     the coset representative $x\in\myFpower{n}/\Cbg$       
     of the coset $y+\Cbg$,
     or equivalently, the syndrome $(\dpr{y}{g_j})_{j=1}^{j=\kcl}$.

\item Bob subtracts the coset representative $x\in\myFpower{n}/\Cbg$ 
      from his code digits, $y-\vare$, and
      corrects the result $y-x-\vare$
      to a codeword $u$ in $C^{\perp}$, where he uses the decoder such that
      $u=y-x$ if $\vare\in\crJ$.

\item Alice uses the coset $(y-x) + C \in C^{\perp}/C$
      and Bob uses $u+C \in C^{\perp}/C$ as the key.

\end{enumerate}

In Step (viii), 
$x\in\myFpower{n}/\Cbg$ means that $x$ is chosen from the
transversal of $\myFpower{n}/\Cbg$ shared by Alice and Bob, 
which may be assumed to be $\crJ$.
In short, by the law of large numbers, 
about $[(1-\pa)\pb+\pa(1-\pb)]\varn$ copies of states are discarded,
about $\pccmp[(1-\pa)(1-\pb)+\pa\pb]\varn$ copies
are used for
transmission of the key with CSS codes, the reliability of
which was evaluated in Section~\ref{sec:exp}, 
and the about $\pc[(1-\pa)(1-\pb)+\pa\pb]\varn$ remaining copies
are used for estimation of the noise level,
which will be explicated in Section~\ref{ss:channel_para}.

In what follows, we will analyze the security of the protocol
under the `individual attack' assumption
that Eve obtains data by an identical measurement on each particle.
Especially, this assumption includes that Eve cannot change her
measurement according to the value of $\rva_i$ or $\rvb_i$.
A measurement is modeled as
a completely positive (CP) instrument whose measurement result
belongs to a finite or countable set (e.g., \cite{HolevoLN,kraus71,hellwig95,kraus,preskillLNbook}).
We also assume that the channel noises $\cN,\cN'$ are tensor products of
identical copies of a CP map.
Namely, we assume a state $\rho\in\Bop(\Hch)$
of each particle suffers a change 
$\rho \mapsto \sum_{i} A_i  \rho A_i^{\dagger}$, and Eve obtains
$i$, or part of it, with probability $\trace A_i^{\dagger}A_i \rho$
as information on this particle.

We remark 
that some quantities such as $\rvz=z$ and the quantum code $(\cQ_{xz},\cR)$
are artifices that have been introduced only to establish the security, 
and are not needed for practice.
For example, in the protocol, 
only half of the decoding operation $\cR$
(the part where a half of the syndrome, viz., 
$(\dpr{x}{g_{i}})_{i=1}^{\kcl}$ in (\ref{eq:eigen4stab})
matters) 
is performed. 
This can be viewed as the decoding for the classical code $\Ccl^{\perp}$
(more precisely, the coset code $y+\Cbg$),
and the decoding error probability of this classical code $\Cbg$,
together with $1-\Expe F_{\rvx\rvz}$
for the corresponding CSS code $\css{\Ccl}{\crJ}$,
has been upper-bounded exponentially in Theorem~\ref{th:main}.

\section{Estimation of Channel Parameters \label{ss:channel_para}}

Roughly speaking, the BB84 protocol consists of
CSS coding and estimation of channel parameters. 
This section explicates 
how the estimation works in the present case of individual attacks.

Since Alice and Bob use the $X$-basis or $Z$-basis at random,
the change suffered by a transmitted state, 
if it is assumed to be a $Z$-basis element $\ket{j}$ initially, is
either $\Eve$ or $\Eve'=\cU^{-1}\Eve\cU$ accordingly as the $Z$-basis ($a_i=b_i=0$) or 
the $X$-basis ($a_i=b_i=1$) is used,
where $\Eve$ represents Eve's action plus the channel noises
for each digit sent,
and $\cU$ denotes the Fourier transform
\begin{equation*} 
\cU(\rho) = U \rho U^{\dagger}
\end{equation*}
with
\begin{equation*} 
U=\dmn^{-1/2}\sum_{j,l\in\myF}\omega^{jl}\ket{j}\bra{l}.
\end{equation*}
Note that the $X$-basis $\{ \ket{j}' \}$ and $Z$-basis $\{ \ket{j} \}$
are related by \[
\ket{j}'=U\ket{j}, \quad j\in\myF. 
\]

We use the following well-known one-to-one map of Choi~\cite{choi75}
between the CP maps on $\Bop(\Hgen)$ and the positive semi-definite
operators in $\Bop(\Hgen\tnsr\Hgen)$:
\begin{equation}\label{eq:choi}
\hMsub{n}{\CPex}= [\Id \tnsr \CPex](\ket{\Psi} \bra{\Psi}),
\end{equation}
where $\Id$ is the identity map on $\Bop(\Hgen)$, 
and $\ket{\Psi}$ is a maximally entangled state given by
\[
\ket{\Psi}=\frac{1}{\sqrt{\Dgen}} \sum_{\varii \in \cB} \ket{\varii}
\tnsr \ket{\varii}
\]
with some orthonormal basis $\cB=\{ \ket{\varii} \}$ of $\Hgen$.
Choi introduced $\dmn^{n}\hMsub{n}{\CPex}$ in matrix form to yield fundamentals
of CP maps.

In the present case, we assume $\ket{\varii} =\ket{\varii_1}\tnsr \cdots \tnsr\ket{\varii_n}$, $\varii=(\varii_1,\dots,\varii_n)\in\myFpower{n}$,
and let
\begin{equation}\label{eq:Phix}
\ket{\Psi_{y}} = 
\frac{1}{\sqrt{\dmn^n}} \sum_{\vara\in\myFpower{n}} \ket{\vara} \tnsr \Ebe_y \ket{\vara}, \quad y \in \cX^{n}.
\end{equation}
These $2n$ vectors form an orthonormal basis of 
$\Hch^{\tnsr n}\tnsr\Hch^{\tnsr n}$ (e.g., \cite{werner01}).
Recall that a symplectic code has a collection of subspaces $\{ \codesubs{\vars} \}$ and recovery operators $\cR_{\vars}$ for each $\vars$, 
where $\vars$ corresponds to the syndrome and has been written as $xz$ for CSS codes.
It is known that an $\Ebe_{\Ksp}$-correcting symplectic code 
$(\codesubs{\vars},\cR_{\vars})$,
used on a channel $\Evn: \Bop(\Hgen)\to\Bop(\Hgen)$,
has
entanglement fidelity, averaged over all $\vars$ with equal probabilities, 
not smaller than $\sum_{y\in \Ksp} P_{\Evn}(y)$:
\begin{equation}\label{eq:boundFenav}
\Expe_{\vars} \Fen(\pi_{\codesubs{\vars}}, \cR_{\vars} \Evn) 
\ge \sum_{y\in \Ksp} P_{\Evn}(y), 
\end{equation}
where $P_{\Evn}(x)$ is associated with the channel $\Evn$ via
\begin{equation}\label{eq:PEve}
P_{\Evn}(x) = \bra{\Psi_{x}} \hMsub{n}{\Evn} \ket{\Psi_{x}}, \quad x\in\cX^{n},
\end{equation}
and $\Expe_{\vars}$ is the expectation operator.
This bound is implicit in \cite{GottesmanPreskill01}
as explained in \ref{app:subsidiary}; 
the bound is tight for the largest choice of $\Ksp$~\cite{hamada03f}. 

Our channel to be analyzed has the product form
$\Evn=\cM^n$, 
and hence $P_{\Evn}$ also has the product form
\[
P_{\Evn}= P_{\cM}^{n}.
\]
Here, 
we have assumed Alice and Bob do not use the values of $\rva_i$ ($=\rvb_i$)
for coding, which implies
that $\cM$ can be regarded as the mixture
\[
\cM=(1-r)\Eve+r\Eve'.
\]
Note, especially in the case where $\dmn=2$, 
$P_{\Eve}$ and $P_{\Eve'}$ are related by
\begin{equation}\label{eq:switch}
P_{\Eve'}(s,t) = P_{\Eve}(t,s), \quad s,t\in\myF,
\end{equation}
since $X$ and $Z$ switches with each other by $\cU$.
More generally, we have
\begin{equation} \label{eq:switch3}
P_{\Eve'}(s,t) = P_{\Eve}(t,-s), \ s,t\in\myF, 
\end{equation}
which is proved in \ref{app:subsidiary}. 

The quantity $P_{\Eve}(\vari,\varj)$ is the probability to obtain 
$(\vari,\varj)$
with a measurement 
$\{ \ket{\Psi_{(\vari,\varj)}}\bra{\Psi_{(\vari,\varj)}}
\}_{(\vari,\varj)\in\myFpower{2}}$ on the system in the state $\hMsub{1}{\Eve}$.
However, this seems hard to implement, so that we divide the problem. 
We measure either $\vari$ or $\varj$ per sample of the state $\hMsub{1}{\Eve}$.
To do this, note that
\begin{equation}\label{eq:ZZ}
Z\tnsr Z^{-1} \ket{\Psi_{(\vari,\varj)}} = \omega^{\vari} \ket{\Psi_{(\vari,\varj)}}
\end{equation}
for $(\vari,\varj)\in\myFpower{2}$.
This implies that measuring eigenvalues of $Z \tnsr Z^{-1}$, i.e.,
performing the measurement $\{ \sum_{\varj\in\myF}
\ket{\Psi_{(\vari,\varj)}} \bra{\Psi_{(\vari,\varj)}} \}_{\vari\in\myF}$
in the state $\hMsub{1}{\Eve}$ gives the result $s$ with probability
$\bar{P_{\Eve}}(s)$.
Measuring eigenvalues of $Z \tnsr Z^{-1}$ is still imaginary,
but measuring eigenvalues of $Z \tnsr I$ and then $I \tnsr Z^{-1}$
is completely simulated by sending one of the eigenstates of $Z$ at random
(according to the uniform distribution)
through $\Eve$ and measuring $Z^{-1}$ at the receiver's end,
and $\bar{\Pev}(\vari)$ equals the probability that the difference 
$\vara-\varaa$
between the sent digit $\vara$ and the received one $\varaa$ is $\vari$. 
For a natural estimate of $\bar{\Pev}(\vari)$ needed in the BB84 protocol,
we use the relative frequency of the appearances of $\vari\in\myF$
in the sequence of the observed differences $\vara_i-\varaa_i$.
In words,
we use the type $\sP_{\rvu}$ of $\sU$ for the estimate of $\bar{\Pev}$,
where the random variable $\rvu$ is the sequence of the differences
$\vara_i-\varaa_i$ and we use only the digits $\vara_i$ and $\varaa_i$
accompanied by $(\rva_i,\rvb_i,\rvc_i)=(0,0,1)$. 
Noticing (\ref{eq:switch3}), we use 
the similar estimates, say, $\sP_{\rvw}$, for $\dbar{\Pev}$,
which is obtained 
from the sequence $\rvw$ of the differences $\varaa_i-\vara_i$
of those $\vara_i$ and $\varaa_i$
accompanied by $(\rva_i,\rvb_i,\rvc_i)=(1,1,1)$. 

\section{Security of the Bennett-Brassard 1984 Protocol \label{sec:security}}

In this section, finally, we will establish the security of the BB84 protocol 
for high rates using Theorem~\ref{th:main}.
This should be done in terms of 
the random variables involved with the protocol, namely,
Alice's sent digits
$\rvetaA=(\rvetaA_{1}, \dots, \rvetaA_{m})$,
Bob's received digits
$\rvetaB=(\rvetaB_{1}, \dots, \rvetaB_{m})$,
$C,\rvx,\rvv, \rva, \rvb, \rvc,\rve$, and $\rvt$ defined below.

In the BB84 protocol, we should consider the
possibility 
of Eve's obtaining knowledges
on the key from the data sent through the public channel,
i.e., $\rvx$, $C$, $\rva$, $\rvb$ and $\rvc$ and the non-code digits
used for the noise estimation 
(in our scheme, $\Gamma$ is determined from $C$,
so that it need not be sent).
For the purpose of analysis, we convert $(\rva,\rvb)$
into $(\rva,\rvd=\rvb-\rva)$, where we regard
$\rva,\rvb$ and $\rvd=(\rvd_1,\dots,\rvd_m)$ as vectors over 
$\mymathbb{F}_{2}$.
Let $\rvaa$ denote
the subsequence $\rva_{\rvt}$ of $\rva$ (\ref{app:nom}),
where $\rvt=\{ i \mid \mbox{$\rvc_i=0$ and $\rvd_i=0$} \}$,
the set of the positions of the code digits 
(with the one element thrown away if $\dmn=2$ and $n=\crd{\rvt}$ 
is initially odd);
let $\rvaaa$ denote the subsequence $\rva_{\rvt\cmple}$
where $\rvt\cmple = \{ 1, \dots, m\} \setminus \rvt$;
we let $\rvestA$ [$\rvestB$] denote the string of publicly announced non-code (estimation) digits of Alice [Bob], 
which is a subsequence of $\rvetaA$ [$\rvetaB$].
Denote the 7-tuple of random variables $(C, \rvaaa, \rvd, \rvc,\rvt,\rvestA,\rvestB)$ by $\rvs$. 
One criterion for security that takes $\rvs$ into account is
$I(\rvv;\rve\rvx\rvaa|\rvs=\rls) \approx 0$ for (almost) every 
definite value of $\rvs=\rls$.
The rationale hereof is that we should evaluate the security
for any definite values of as many parameters as possible.
To show that our scheme fulfills this criterion, we modify the argument
in Section~\ref{ss:principle_sch} as follows.

The argument in Section~\ref{ss:principle_sch} 
is applicable to the above protocol if 
we add
the conditioning on $\rvaa$ and $\rvs$ 
to the mutual informations $I$. 
Specifically,
we begin with $I(\rvv;\rve|\rvx=x,\rvz=z,\rvaa=a',\rvs=s) \le \Senew{xz}{a',s}$
instead of (\ref{eq:Sch1}).
Note that what we have evaluated above is the fidelity 
$\Expe_{\rvx\rvz\rvaa}F_{\rvx\rvz,\rvaa,s}$ (and the decoding error probability for key transmission) of the codes used on the channel $\cM^{\tnsr n}$,
where $\Senew{xz}{a',s}$ and $F_{xz,a',s}$ are the obvious replacements
for $\Se{x}{z}$ and $F_{xz}$ with conditioning on $\rvaa=a'$ and $\rvs=s$, and
$\Expe_{\rvy}$ denotes the expectation operator with respect to a random variable $\rvy$.
Then, in this case, we can replace (\ref{eq:Sch5}) with
\begin{equation}
I(\rvv;\rve|\rvx\rvz\rvaa,\rvs=\rls)
\le 2\dmn^{-nE+o(n)} [n(E  +  R) -o(n)] \label{eq:Sch6}
\end{equation}
using the bound
$1-\Expe F_{\rvx\rvz\rvaa,s} \le \dmn^{-n E(R,\bar{P},\dbar{P}) + o(n)}$
in Theorem~\ref{th:main}.
From the chain rule of mutual information~\cite{CsiszarKoerner,CoverThomas}, 
we have
\begin{eqnarray*}
\lefteqn{I(\rvv;\rve\rvx\rvz\rvaa | \rvs=\rls)}\\
&=& I(\rvv;\rvx\rvz\rvaa|\rvs=\rls) + I(\rvv;\rve|\rvx\rvz\rvaa,\rvs=\rls) ,
\end{eqnarray*}
where $I(\rvv;\rvx\rvz\rvaa|\rvs=\rls)=0$ due to the mutual independence of $\rvv$ from $\rvx,\rvz,\rvaa$ given $\rvs=\rls$,
and hence, 
$I(\rvv;\rve\rvx\rvaa|\rvs=\rls) \le I(\rvv;\rve\rvx\rvz\rvaa|\rvs=\rls) = I(\rvv;\rve|\rvx\rvz\rvaa,\rvs=\rls)$.
Combining this with (\ref{eq:Sch6}), we obtain
\begin{equation}
I(\rvv;\rve\rvx\rvaa|\rvs=\rls)
\le 2\dmn^{-nE+o(n)} [n(E  +  R) -o(n)]. \label{eq:Sch7}
\end{equation}
Note that $n$ is also a random variable, which is a function of 
$m$ and $\rvs=\rls$.

Now it is time to clarify the meaning of what is stated in Step (vii) 
of the BB84 protocol in Section~\ref{sec:BB84}.
Recall our assumption 
$\pa,\pb,\pc\in (0,1)$ and (\ref{eq:code2whole}), which imply
\[
0 < \vra < 1,
\]
as well as that the channel $\cM=(1-r)\cA+r\cA'$ stands for Eve's action,
which implies
$\bar{P_{\cM}} = (1-r) \bar{P_{\cA}}+ r \flp{\dbar{P_{\cA}}}$
and $\dbar{P_{\cM}} = (1-r) \dbar{P_{\cA}} + r \bar{P_{\cA}}$,
where 
the operation $\mymathsf{f}$ on probability distributions is defined by
\[
\flp{q}(t)=q(-t), \quad t\in\myF.
\]

Let Alice and Bob choose
a moderate number $E>0$ as a wanted speed
of convergence of the amount of the possible information leakage 
$I(\rvv;\rve\rvx\rvaa|\rvs=\rls)$ as well as a sufficiently
small positive constant $\vep$.
They use the estimate $\sP_{\rvu}$ of $\bar{P_{\cA}}$ and the estimate
$\sP_{\rvw}$  of $\dbar{P_{\cA}}$ in Section~\ref{ss:channel_para}.
Let $\cG$ consists of triples $(\alpha,p,q)$, where 
$0 \le \alpha \le 1$, and $p, q$ are distributions on $\myF$. 
With a triple $(\alpha,p,q)\in\cG$,
we associate a probability distribution on
$\{0,1 \} \times \myF$, which we denote by
$\pxp{\alpha,p,q}$ and specify by 
$\pxp{\alpha,p,q}(0,x)=(1-\alpha)p(x)$ and
$\pxp{\alpha,p,q}(1,x)=\alpha q(x)$, $x\in\myF$.
Let $\vma$ [$\vmb$] denotes the number of samples used for the 
estimation of $\bar{P_{\Eve}}$ [$\dbar{P_{\Eve}}$],
and put $\nu=\vma+\vmb$.

In Step (vii), 
they choose a rate $R$ such that $E(R,(1-\alpha)p+\alpha\flp{q},(1-\alpha)q+\alpha p)\ge E$
for any triple $(\alpha,p,q)\in\cG$ such that
$\| \pxp{\alpha,p,q} - \pxp{\vmb/\nu,\sP_{\rvu},\sP_{\rvw}} \|_1 \le \vep$,
and a code of rate 
$R$ and fidelity
not smaller than $1- {\dmn}^{ - n E(R,\bar{P_{\cL}},\dbar{P_{\cL}}) + o(n) }$
for any channel $\cL$,
the existence of which is ensured by Theorem~\ref{th:main}.
[The function $o(n)$ is explicitly given in Remark~4 to Theorem~\ref{th:main}.]

For simplicity, we restrict our attention to the almost sure event
where $\nu/m \to [(1-\pa)(1-\pb) + \pa\pb]\pc>0$
as $m\to\infty$, which directly follows
from the strong law of large numbers applied to 
$(\rvd_i,\rvc_i)$, $i=1,2,\dots$ (e.g., \cite{billingsleyPM}).
For any $m$, if $\| \pxp{r,\bar{P_{\cA}},\dbar{P_{\cA}}} -
\pxp{\vmb/\nu,\sP_{\rvu},\sP_{\rvw}}
\|_1 \le \vep$, then $E(R,\bar{P_{\cM}},\dbar{P_{\cM}}) \ge E$ as desired.
Owing to (\ref{eq:prob_type}), the probability (conditioned on specific values of $\rvd,\rvc$) of the event 
of estimation failure where 
$\| \pxp{r,\bar{P_{\cA}},\dbar{P_{\cA}}} - 
\pxp{\vmb/\nu,\sP_{\rvu},\sP_{\rvw}} \|_1 > \vep$
is upper-bounded by
\begin{equation}
\dmn^{ - \nu \min_{ Q :\, \big\| 
Q - \pxp{r,\bar{P_{\cA}},\dbar{P_{\cA}}} \big\|_1 \ge \vep }
D\big(Q||\pxp{r,\bar{P_{\Eve}},\dbar{P_{\Eve}}}\big)  +o(m) },
\end{equation}
and this goes to zero with probability one in our almost sure event.

Hence, the above version of the BB84 protocol is secure 
in the sense that with Eve's attack
modeled as a tensor product form of identical copies of
a CP instrument, for any such instrument, either
`the mutual information between the key and the eavesdropper's obtained data,
together with the decoding error probability for the key transmission, 
is upper-bounded by ${\dmn}^{ - n E + o(n)}$,
where $E$ is positive'
or 
`the probability that the detection of eavesdroppers fails
is exponentially close to zero'. 
Especially, reliable and secure key transmission is possible with this protocol
at any rate below 
\begin{eqnarray}
\mbox{}\!\!\!\!\! 
\lefteqn{\pccmp(1-\pa-\pb+2\pa\pb)}\nonumber\\
&&\cdot[
1-2 \max\{ H(
(1-r)\bar{P_{\Eve}}+
r\flp{\dbar{P_{\Eve}}}),
H(
(1-r)\dbar{P_{\Eve}}+
r\bar{P_{\Eve}}) \}
)],
\label{eq:Rqkd}
\end{eqnarray}
where the rate indicates the ratio of the length of the key to
the number of uses of the channel, 
rather than to the code length of the incorporated CSS code.

\section{Discussions \label{ss:discussions}}

The achievability of the rate $[1-2 \hmo(\erx+\erz)]/4$,
where $\erx=\dbar{P_{\Eve}}(1),\erz=\bar{P_{\Eve}}(1)\le 1/2$
may be implicit in \cite{ShorPreskill00}
though their error rates may
differ from our $\erx$ and $\erz$.
This bound can be understood to be obtained by using the exponent
$E_{\rm GV}(R,P_{\cM}) = \min_{ 1-2\hmo(\bar{Q}(1)+\dbar{Q}(1))
\le R\ {\rm or}\ \bar{Q}(1)+\dbar{Q}(1) \ge 1} D(Q||P_{\cM})$
in place of $E(R,\bar{P_{\cM}},\dbar{P_{\cM}})$ 
of Theorem~\ref{th:main}. 
Specifically, this follows
from the Gilbert-Varshamov bound for CSS codes~\cite{CalderbankShor96}
and Sanov's theorem in large deviation theory (e.g., \cite{DemboZeitouni,CoverThomas}) or (\ref{eq:prob_type}).
[For the present purpose, we need only
the upper bound on the probability in question, so that the half of
Sanov's theorem, viz., (\ref{eq:prob_type}), is enough.]
Shor and Preskill~\cite{ShorPreskill00} also mentioned a higher rate, which corresponds to 
$[1-2 \hmo((\erx+\erz)/2)]/4$,
i.e., (\ref{eq:Rqkd}) with $\pa=\pb=\pc=1/2$ or
\begin{equation}\label{eq:Rmixture}	
[1-2H(P_{\cM})]/4 \quad (\dmn=2).
\end{equation} 
This rate is established by Theorem~\ref{th:main} (Section~\ref{sec:exp})
rigorously.
Another achievable rate is presented in \ref{app:newdec}.
Several other achievable rates (or tolerable error rates) 
have been mentioned in the literature
(e.g., \cite[Eq.~(38)]{GottesmanPreskill01}, \cite{Lo01,LoChauA00})
without details on their code structures.

\section{Conclusion \label{sec:conc}}

In summary, we have established
achievable rates in the BB84 protocol.
This improves the one based on the Gilbert-Varshamov bound for CSS codes,
which may be implicit in Shor and Preskill's security proof.
Specifically, in this paper proved was the existence of a version of
the BB84 protocol with
exponential convergence of the mutual information between
Alice and Eve to zero for any rate below the number in (\ref{eq:Rqkd}).
Several issues lacking 
in the literature were pointed out and resolved
(cf.\ criticisms of Yuen~\cite[Appendix A]{yuen03} on other security proofs).
Namely, the existence of CSS codes robust against
fluctuations of channel parameters was proved, and
the decoding error probability for key transmission, together with
the mutual information, was shown to decrease exponentially.
Especially, it was proved that codes of `balanced' weight spectra
(Corollary~\ref{coro:goodcode})
achieve the coding rate $1-2 \hmo((\erx+\erz)/2)$
for $\dmn=2$, where
$\erx=\dbar{P_{\Eve}}(1),\erz=\bar{P_{\Eve}}(1)$.
A proof of the security of a BB84-type protocol
for joint attacks is given in \ref{app:joint}.

In a seemingly less practical but theoretically interesting setting where Eve's attack is known to Alice and Bob beforehand,
the optimum rate has recently been obtained in \cite{devetak03}.

\section*{Acknowledgment}

The author appreciates a comment of Masahito Hayashi on an earlier version  
of this paper that the security proof should 
extend to the case of joint attacks
as well as valuable discussions with him.
The author is grateful to Hiroshi Imai for support.

\appendix

\mysectionapp{Proofs of Subsidiary Results \label{app:subsidiary}}

\mysubsectionapp{Proof of the fidelity bound (\ref{eq:boundFenav})}

The bound directly
follows from the argument in the two paragraphs containing Eqs.~(18)--(24)
of \cite[Section~III-B]{GottesmanPreskill01} for $\dmn=2$.
The entanglement distillation protocol they used is the same as
Shor and Preskill's~\cite{ShorPreskill00} and can be interpreted as
follows for our purposes.
Given a bipartite state $\hMsubbar{n}{\Evn}= [\Id \tnsr \Evn](\ket{\bar{\Psi}} \bra{\bar{\Psi}})$, 
where $\ket{\bar{\Psi}} = \dmn^{-n/2} \sum_{\vars,y}\ket{\bar{\vars,y}}\tnsr
\ket{\bar{\vars,y}}$, where $\{ \ket{\bar{\xi,y}} \}_y$ is an orthonormal basis of $\codesubs{\vars}$. 
Alice performs the local measurement $\{ \Pi_{\vars} \}$
on the first half of the system, where
$\Pi_{\vars}$ denotes the projection onto the code space $\codesubs{\vars}$, 
and Bob performs the recovery operation for the $\Ebe_{\Ksp}$-correcting code
$\codesubs{\vars}$ knowing that Alice's measurement result is $\vars$.
Since Alice obtains each result $\vars$ with the equal probability,
the lower bound of \cite{GottesmanPreskill01} serves as that on
the average entanglement fidelity of the code $(\codesubs{\vars},\cR_{\vars})$
in question.

The bound (\ref{eq:boundFenav}) for $\dmn \ge 2$,
together with its tightness, follows from
the formula for `discrete twirling' (\cite{hamada03t} and references therein)
and the properties of the symplectic codes~\cite{hamada03f}.
It is remarked that a similar bound was given
by the present author~\cite[Lemma~5]{hamada01g};
we can rephrase this bound in terms of the entanglement fidelity $\Fen$
using the relation
\[
K(K+1)^{-1}[1-\Fen(K^{-1}I,\cA)] = 1-\Expe_{\varphi} \bra{\varphi} \cA (\ket{\varphi}\bra{\varphi})
\ket{\varphi},
\]
where $\cA$ is a CP map on $\Bop(H)$ with $\dim H=K$,
and $\Expe_{\varphi}$ denotes the expectation operator with
$\varphi=\ket{\varphi}$ regarded as uniformly distributed over all
unit vectors in $H$~\cite{HHH99t},
though the resulting bound has the form
$1-\Fen' \le (K+1)K^{-1} \sum_{y\in J\cmple} P_{\Evn}(y)$, 
which is 
weaker than (\ref{eq:boundFenav})
by the asymptotically negligible factor of $(K+1)K^{-1}$.

\mysubsectionapp{Proof of (\ref{eq:switch3})}

First, observe, by the definition of $\hMsubnoarg{1}$ in (\ref{eq:choi})
and that of $\ket{\Psi_y}$ in (\ref{eq:Phix}),
that $P_{\cA}(s,t)$ can be written as
\[
P_{\cA}(s,t)=\sum_{i} \left| \dmn^{-1} \trace A_i^{\dagger} X^sZ^t \right|^2, \quad s,t\in\myF
\]
for a CP map $\cA(\sigma)=\sum_{i} A_i \sigma A_i^{\dagger}$.
Then, for $\Eve'=\cU^{-1}\Eve\cU$, we have
\begin{eqnarray*}
P_{\Eve'}(s,t) &=& \sum_i \left| \dmn^{-1} \trace (U^{\dagger}A_i U)^{\dagger} X^sZ^t \right|^2\\
&=& \sum_i \left| \dmn^{-1} \trace A_i^{\dagger} UX^s U^{\dagger}U Z^t U^{\dagger} \right|^2\\
&=& \sum_i \left| \dmn^{-1} \trace A_i^{\dagger} Z^{-s} X^t \right|^2
\end{eqnarray*}
where we used the relations $UXU^{\dagger}=Z^{-1}$ and 
$UZU^{\dagger}=X$ for the last equality.
Since $Z^{-s}X^t$ is the same as $X^tZ^{-s}$ up to a phase factor, 
$\omega^{st}$, by 
the commutation relation $XZ=\omega ZX$ or (\ref{eq:WCR}),
we have $P_{\Eve'}(s,t)=P_{\Eve}(t,-s)$, as promised.

\mysubsectionapp{Proof That $\bar{P}^{n}(\crJ\sqi{n}\prcmple)$ Is the Decoding Error Probability for Key Transmission \label{appsub:epk}}

The probability in question has the form
$[\tdP_1\cdots \tdP_n](T)$, where $\tdP_i$ are probability distributions 
on $\myF$ and $T\subset\myFpower{n}$ [in the present case, $\tdP_i$ are
identically equal to $\bar{P}$],
while the $i$\/th transmitted digit suffers the probabilistic change
described by a channel matrix, say, $Q_i(y_i |x_i)$
with $\tdP_i(z_i)=\dmn^{-1}\sum_{x_i\in\myF} Q_i(x_i-z_i|x_i)$ as already 
argued in
Section~\ref{ss:channel_para}.
Putting $q_i(z_i|x_i)=Q_i(x_i-z_i|x_i)$,
$[q_1\dots q_n](z_1,\dots,z_n | x_1,\dots,x_n) = q_1(z_1|x_1)\cdots q_n(z_n|x_n)$,
and recalling the decoding procedure in Steps (viii)--(x) 
of the protocol,
we see the decoding error probability is given by 
$\dmn^{-n}\sum_{x\in\myFpower{n}} [q_1 \cdots q_n](T|x)
= [\tdP_1\cdots \tdP_n](T)$, as desired.

\mysectionapp{Minimum Conditional Entropy Decoding \label{app:newdec}}

In this appendix, 
a decoding strategy for CSS codes in the BB84 protocol
that results in an improvement on the achievable rate,
especially when $r=1/2$,
is proposed.

Define $\mua_{\varn}$ and $\mub_{\varn}$ by
$
\mua_{\varn}=\crd{\{ i \mid 1\le i\le\varn,\ (\rva_i,\rvb_i,\rvc_i)=(0,0,0) \}}
$
and
$
\mub_{\varn}=\crd{\{ i \mid 1\le i\le\varn,\ (\rva_i,\rvb_i,\rvc_i)=(1,1,0) \}}
$,
where $\varn$ is the number of the whole sent digits.
In the proposed scheme, 
Alice and Bob use $\min\{ \mua_{\varn}, \mub_{\varn} \}$ digits with 
$(\rva_i,\rvb_i,\rvc_i)=(0,0,0)$ and the same number of  digits with 
$(\rva_i,\rvb_i,\rvc_i)=(1,1,0)$ for CSS coding
discarding excessive digits if they exist. 
If $r=1/2$, the loss of digits in this process is small
by the strong law of large numbers. 

In the conventional decoding schemes for CSS codes in the BB84
protocol~\cite{ShorPreskill00,GottesmanPreskill01,GottesmanLLP02}
or that in Section~\ref{sec:security},
Bob does not use the information as to whether 
$\rva_i=\rvb_i=0$ or $\rva_i=\rvb_i=1$ has occurred; 
he considers the channel as the mixture of
$\Eve$ and $\Eve'=\cU^{-1}\Eve\cU$.
To improve on the achievable rates in (\ref{eq:Rqkd}) for $r=1/2$,
we employ a decoding strategy
that uses the information on $\rva_i$ ($=\rvb_i$), 
minimum conditional entropy decoding, so to speak.
Specifically, we associate each word $\xa\xb$,
where $\xa\xb$ denotes 
the concatenation of $\xa\in\myFpower{\nua}$ and
$\xb\in\myFpower{\nub}$, and $\xa$ [$\xb$] is composed of the digits for which 
$\rva_i=0$ [$\rva_i=1$], with the conditional entropy 
\begin{equation}
\Hcndn{\xa}{\xb}=
\frac{H(\sP_{\xa}) + H(\sP_{\xb})}{2}, 
\end{equation}
and {\em choose a word that minimizes
the conditional entropy $\Hcndnoarg$ in each coset in
$\myFpower{n}/\Ccl^{\perp}$ 
to obtain a transversal $\crJ$.}
The quantity $\Hcndn{\xa}{\xb}$ can be written solely with
$\sP_{\xa}$ and $\sP_{\xb}$, so that we will occasionally denote
$\Hcndn{\xa}{\xb}$ by
$\Hcndn{\sP_{\xa}}{\sP_{\xb}}$.

\begin{theorem} \label{th:main2a}
Let a number $0 \le R \le 1$ be given.
There exists a sequence of pairs 
$\{ (C\sqi{\nua}, \crJ\sqi{\nua}) \}_{\nua\in\SNN}$, each consisting
of a self-orthogonal code
$C\sqi{\nua} \subset \myFpower{2\nua}$ with
$2\nua -2 \dim C\sqi{\nua} \ge 2\nua R$
and a set $\crJ\sqi{\nua}$ of coset representatives 
of $\myFpower{2\nua}/C\sqi{\nua}^{\perp}$ such that
for any pair of probability distributions $\Pa$ and $\Pb$ on $\cX$, 
\[
\Pa^{\nua}\Pb^{\nub}(\Kof{\crJ\sqi{\nua}'}\cmple) \le
\bar{\Pa}^{\nua}\bar{\Pb}^{\nub}(\crJ\sqi{\nua}\prcmple)+
\dbar{\Pa}^{\nua}\dbar{\Pb}^{\nub}(\crJ\sqi{\nua}\prcmple)
\le \dmn^{-2\nua E_{\rm c}(R,\Pa,\Pb) + o(\nua)},
\] 
where
\[
\crJ'\sqi{\nua}=\crJ\sqi{\nua}+C\sqi{\nua},
\]
\[
E_{\rm c}(R,\Pa,\Pb) = \min \{ \Estar(R,\bar{\Pa},\bar{\Pb}), \Estar(R,\dbar{\Pa},\dbar{\Pb}) \},
\]
\begin{eqnarray*}
\Estar(R,\vpa,\vpb) &=&
\min_{\Qa,\Qb} \big[ D(\Qa||\vpa) + D(\Qb||\vpb) \\
&&\ \ \ \ \ \ \ \ \ \ \mbox{}+ 
|1- 2 \Hcndr{\Qa}{\Qb} - R  |^+ \big]/2,
\end{eqnarray*}
and the minimization with respect to $(\Qa,\Qb)$ is taken 
over all pairs of probability distributions on $\myF$.
\end{theorem}
The proof is similar to that of Theorem~\ref{th:main}.
In this case, we pair up digits in a sequence
$xy=(x_1,\dots,x_{\nu},y_1,\dots,y_{\nu})$
as $(x_1,y_1),\dots,(x_{\nu},y_{\nu})$ to regard it
as a sequence from $\cX^{\nu}$, $\cX=\myF\times\myF$.
Then, to evaluate the fidelity of the codes, 
we use the existence proof of `balanced' codes 
in Section~\ref{sec:exp}, which is clearly valid
if we use types in $\cP_{\nu}(\cX)$
in place of types in $\cP_{n}(\myF)$, and the similarly modified
permutation argument for sequences in $\cX^{\nu}$. 
By this theorem with $P_0=P_{\Eve}$ and $P_1=P_{\cU^{-1}\Eve\cU}$,
the rate $(1-\pc)[1-H(\bar{\Eve})-H(\dbar{\Eve})]/2$ is
achievable with the BB84 protocol.
The result extends to an arbitrary rational $r$;
for example, for $r=1/3$, we can use types in $\cP_{\nu}(\myFpower{3})$.

\mysectionapp{Security against Joint Attacks \label{app:joint}}

In this appendix,
we will prove the security of the following modified
BB84 protocol against any joint attack
through this paper's approach. Especially, 
an exponential upper bound on the information leakage
to Eve, which holds for finite $m$ and $n$, will be established.
This modification to the protocol is essentially due to \cite{LoChauA00},
and its main idea 
is as follows. In the protocol,
about $\pa\pb m$ digits with $\rva_i = \rvb_i=1$ are used for estimation of 
the level of errors caused by the Weyl unitary $Z$,
the same number of randomly chosen 
digits with $\rva_i = \rvb_i=0$ are used for estimation 
of those caused by $X$,
and the about $[(1-\pa)(1-\pb)-\pa\pb] m$ 
remaining digits with $\rva_i = \rvb_i=0$ are used for CSS coding.
In this paper, 
we assume that the parameters $\pa=\Pr\{ \rva_i = 1 \},\pb=\Pr\{ \rvb_i = 1 \}\in (0,1/2)$ 
are independent of $m$
in order that the law of large numbers (or any other refined law such as Sanov's theorem) is applicable to $\{ (\rva_i,\rvb_i) \}$;
in \cite{LoChauA00}, 
it is assumed $\pa,\pb$ depend on $m$ so that $\vra$ in (\ref{eq:code2whole})
goes to $0$ as $m$ goes to infinity 
(seemingly only for the purpose of analysis of security);
Hayashi~\cite{hayashi03pc}
described an idea for a possible proof of security of this protocol
using the codes in Theorem~\ref{th:main} 
(in fact, the modification for $\dmn=2$ in Section~\ref{ss:exp2}) 
for small enough $\vra$.

Let $\cS_n$ be the symmetric group on $\{ 1,\dots, n\}$ as before. 
For the proof for joint attacks, 
we should be more specific about the expression
of the key. Given a self-orthogonal code $C$, the key, which is actually
a string of $k=\rlcl-2\kappa$ digits, is encoded into $C^{\perp}/C$.
The encoding map, $f_{C,h_1,\dots,h_k}$, can be given as
$f_{C,h_1,\dots,h_k}:
(\sigma_1,\dots,\sigma_{k}) \mapsto C+\sigma_1 h_{1}+\dots+\sigma_k h_{k}$,
where $\{ h_{1},\dots,h_{k} \}$, together with a basis of $C$,
gives a basis of $C^{\perp}$. 
Thus, Alice and Bob specify
their cryptographic code by
$(g_1,\dots, g_{\kappa}; h_{1},\dots,h_{k}; \Gamma)$;
in this appendix, we always assume $g_1,\dots,g_{\kappa}$ form a basis of $C$.
We use the syndromes 
$\rvx',\rvz'\in\myFpower{\kappa}$ for the code $\Cbg$ 
and the coset representatives $\rvx,\rvz$ for $\myFpower{n}/\Cbg$
interchangeably since they
are in one-to-one correspondence with each other
once the generator $g_1,\dots, g_{\kappa}$ of $C$ is fixed:
$\rvx H\transp = \rvx'$, 
where $H\transp=[g_1\transp \cdots g_{\kappa}\transp]$.
In places where we want to distinguish a random variable
from its realization, we use the sanserif or bold font for the former
and the italic font for the latter as in the text.

\begin{center}
{\bf Modified BB84 protocol}
\end{center}
\begin{enumerate}
\item   The sender, Alice, and the receiver, Bob, do Steps (ii)--(iv) 
for each $i=1,\dots, \varn$.

\item   Alice chooses a random bit $\rva_i$.
        She prepares her system in one state that is chosen uniformly randomly from the $Z$-basis
	if $\rva_i$ is $0$, or
        in one from the $X$-basis if $\rva_i$ is $1$.

\item Alice sends the prepared state to Bob.

\item Bob chooses another random bit $\rvb_i$,
      and receives the state, 
      performs the $Z$-basis measurement
      if $\rvb_i$ is $0$, or 
      $X$-basis measurement if $\rvb_i$ is $1$.
	
\item Alice and Bob announce $\rva=(\rva_1,\dots,\rva_{\varn})$ and $\rvb=(\rvb_1,\dots,\rvb_{\varn})$, respectively.

\item Alice and Bob discards any results where $\rva_i \ne \rvb_i$.
Let $\rvtsift=\{ i \mid \rva_i = \rvb_i \}$ (the remaining places) and
$\rvms=\crd{\rvtsift}$.
[In the case where $\dmn=2$,
it is assumed that $\rvms$ is even; 
if not, they 
disregard another place chosen randomly from $\{ i \mid \rva_i = \rvb_i \}$.]
Put $\rvcl \defeq \rvms-2\crd{\{ i \mid \rva_i = \rvb_i = 1 \}}$.
If $\rvcl \le 0$ or $\rvcl=\rvms$, they abort the protocol.
To divide $\rvtsift=\{ i \mid \rva_i = \rvb_i \}$ into two parts,
i.e., that for CSS coding $\rvt$, and that for estimation for the noise level
$\rvtsift \setminus \rvt$, they do the following.
From $\{ i \mid \rva_i = \rvb_i = 0 \}$, Alice randomly chooses
(according to the uniform distribution over all possible choices)
$(\rvms-\rvcl)/2=\crd{\{ i \mid \rva_i = \rvb_i = 1 \}}$ 
places where digits are to be used for estimation of the level
of errors caused by the Weyl unitary $X$, 
and tells the choice to Bob. 
The set of the remaining $\rvcl$ places with $\rva_i = \rvb_i = 0$
constitute $\rvt$.
The digits with $\rva_i = \rvb_i = 1$ will also be used for noise estimation.

\item Alice and Bob announce the values of their estimation digits
thus chosen (which will be $\rvestA$ and $\rvestB$ below) and from these, 
estimate the noise level, and decide
on a secure transmission rate, and a CSS code $(g_1,\dots, g_{\kappa}; h_1,\dots,h_k; \crJ)$ to be used.

\item Alice chooses a random permutation $\prm$
from $\cS_{\rvcl}$ according to the uniform distribution,
and tells the choice to Bob. 

\item Alice announces the coset $y+\prm(\Cbg)$, where $y$ ($=w+v+x$) is the string
     consisting of the remaining code digits. In other words,
     she announces the syndrome $\rvx'=(\dpr{y}{\prm(g_j)})_{j=1}^{j=\kcl}$,
     which is in one-to-one correspondence with 
     the coset representative $x\in\myFpower{n}/\prm(\Cbg)$       
     of the coset $y+\prm(\Cbg)$.

\item Bob subtracts the coset representative $x\in\myFpower{n}/\prm(\Cbg)$ 
      from his code digits, $y-\vare$, and
      corrects the result $y-x-\vare$
      to a codeword $u$ in $\prm(C)^{\perp}$, where he uses the decoder such that
      $u=y-x$ if $\vare\in\pi(\crJ)$.

\item Alice uses $\mbm{\sigma}=f_{\prm(C),\prm(h_1),\dots,\prm(h_k)}^{-1}[y-x+\prm(C)]$
      and Bob uses $\mbm{\sigma}'=f_{\prm(C),\prm(h_1),\dots,\prm(h_k)}^{-1}[u+\prm(C)]$ as the key.

\end{enumerate}

Let a TPCP map $\cA: \Bop(\sH^{\tnsr\varn})\to\Bop(\sH^{\tnsr\varn})$ 
represents the whole action of Eve (plus the other environment).
This means that there exists a decomposition (CP instrument) $\{ \cA_i \}_i$
such that $\cA=\sum_{i} \cA_i$, where $\cA_i$ are 
trace-nonincreasing CP maps, and when the initial state of the system of
the whole sent digits is $\rho$,
Eve obtains data $\rve=i$ with probability $\trace \cA_i(\rho)$
leaving the system in state $\cA_i(\rho) /\trace \cA_i(\rho) $.
Here, the decomposition may depend on the other random variables available to Eve.
However, the proof 
relies on the assumption that 
$\cA$ does not depend on $\rva$, $\rvb$, which is needed to use Lemma~\ref{lem:sampling} below.
Recalling the interpretation of
the $Z\tnsr Z^{-1}$ measurement in Section~\ref{ss:channel_para}
and using the $\bar{U}\tnsr U$-invariance of $\ket{\Psi}$,
where $U=\dmn^{-1/2}\sum_{j,l\in\myF}\omega^{jl}\ket{j}\bra{l}$ and $\bar{U}=U^{-1}$, and the relation
$X \tnsr X \ket{\Psi_{(\vari,\varj)}} = \omega^{\varj} \ket{\Psi_{(\vari,\varj)}}$ in addition to (\ref{eq:ZZ}),
we notice that 
Alice's sent digits
$\rvetaA=(\rvetaA_{1}, \dots, \rvetaA_{m})$
and Bob's received digits
$\rvetaB=(\rvetaB_{1}, \dots, \rvetaB_{m})$
are mathematically
equivalent to the results of the following fictional measurements.
We imagine that Alice and Bob have a bipartite system
in state $\hMsub{\varn}{\cA}$, and observe 
$O^{(i)}_{a_i}$, $i =1,\dots,m$,
and $O^{(i)}_{b_i}$, $i =1,\dots,m $, respectively,
where
$O^{(i)}_{a_i}=I^{\tnsr (i-1)}  \tnsr O_{a_i} \tnsr I^{\tnsr (m-i)} \in \Bop(\sH^{\tnsr m})$.
Here,
$O_0$ is the `observable' $Z$ to distinguish the eigenvalues of $Z$
(more precisely, the $Z$-basis measurement $\{ \ket{i}\bra{i} \}_{i=0}^{\dmn-1}$),
and $O_1$ denotes $X$, i.e, $\{ \ket{i}'\bra{i}' \}_{i=0}^{\dmn-1}$. Then,
$\rvetaA$ and $\rvetaB$ are the same as the sequence of
the measurement results of Alice and that of Bob, respectively, for $\dmn=2$
(and this is true if each digit $t\in\myF$ of $\rvetaA$ with $\rva_i=1$
is replaced by $-t$ for $\dmn>2$).
Moreover, we can relate $\rvetaA$ and $\rvetaB$ to
the classical random variables
$(\rvxi_i,\rvzeta_i)$, $i=1,\dots,m$,
which are drawn according to $P_{\cA}$
defined by (\ref{eq:PEve}) as follows.
We have $\rvxi_{\rvt_0}=\rvetaA_{\rvt_0}-\rvetaB_{\rvt_0}$
for the subsequence $\rvxi_{\rvt_0}$ of 
$\rvxi=\rvxi_1\cdots\rvxi_{m}$ (\ref{app:nom}), 
where $\rvt_0=\{ i \mid \rva_i=\rvb_i=0 \}$,
and 
$\rvzeta_{\rvt_1}=\rvetaB_{\rvt_1}-\rvetaA_{\rvt_1}$,
where $\rvt_1=\{ i \mid \rva_i=\rvb_i=1 \}$.

In what follows, we evaluate the fidelity, $F_{\rlss}$ defined below, 
of the symplectic code underlying the protocol,
which is, in essence, the CSS code $(g_1,\dots,g_{\kappa};h_1,\dots,h_k;\crJ)$.
As before, the fidelity can be written in terms of 
the classical random variables
$\rvxi,\rvzeta$.
[In fact, the underlying code is the combined system of this CSS code
$\css{C}{\crJ}$
and a trivial symplectic code, which conveys no information
(i.e., protects only a one-dimensional subspace
of $\sH^{\tnsr (m-n)}$), where each code does its job independently.
The trivial code is 
the collection of simultaneous eigenspaces of
$\{ O^{(i)'}_{a_i} \mid i\in \{1,\dots,m \} \setminus \rvt \}$,
where $O^{(i)'}_{a_i}\in \Bop(\sH^{\tnsr (m-n)})$ is obtained from 
$O^{(i)}_{a_i}=I^{\tnsr (i-1)}  \tnsr O_{a_i} \tnsr I^{\tnsr (m-i)}\in \Bop(\sH^{\tnsr m})$
by neglecting $I$'s on the systems for $\rvt$.
The combined code is 
an $N_{\Kof{\crJ}\times \myFpower{2(m-n)}}$-correcting symplectic code.
Here, the appropriate permutation on $\{ 1,\dots, m \}$ is to be
understood.]

Let $\rvestA$ [$\rvestB$] denote the string of publicly announced estimation digits of Alice [Bob], which is a subsequence of $\rvetaA$ [$\rvetaB$];
assume, say, the first half of $\rvestA$ [$\rvestB$] consists of the digits 
accompanied by $\rva_i=\rvb_i=0$ and the latter half is for $\rva_i=\rvb_i=1$.
Recall $\rvt \subset \{ 1,\dots,m \}$ denotes
the set of the positions of the code digits.
Eve can have access to
$\rva$, $\rvb$,
$\rvx'$,
$\rvss=(\rvtsift,\rvms,\rvcl)$,
$\rvestA,\rvestB$,
$\rvt$, $\rvk$, 
$\rvcode=(g_1,\dots,g_{\kappa};h_1,\dots,h_k;\Gamma)$ and
$\rvpi$. 
In what follows, 
with $\varn>0$ and the realization $\rvss=\rlss$ arbitrarily fixed,
we will upper-bound 
$I(\mbm{\sigma};\rve\rvx'\rvestA\rvestB\rvpi\rvt\rva\rvb\rvcode|\rvk,\rvss=\rlss)$
and the probability of key disagreement 
$\Pr\{ \mbm{\sigma} \ne \mbm{\sigma}' | \rvss=\rlss \}$
simultaneously.

In step (vii), Alice and Bob choose the code in the following manner.
With a sufficiently small constant $\gamma>0$ chosen beforehand, 
they set 
\begin{equation}\label{eq:chosenrate}
R(n,\rlms,\rvestA,\rvestB)=1-2\max\{H(\sP_{\rvxiest}),H(\sP_{\rvzetaest})\}-2\gamma,
\end{equation}	
where $\rvxiest$ and $\rvzetaest$
represent the first half of $\rvestA-\rvestB$
and the second half of $\rvestB-\rvestA$, respectively,
calculate the minimum $k$ of the possible code size
$k'$ with $k'/\rlcl \ge R(n,\rlms,\rlestA,\rlestB)$,
set $\rvk=k$,
and choose a code 
$(g_1,\dots,g_{\kappa}; h_1,\dots,h_k ;\crJ)$
from the shared list of codes satisfying the property of
Corollary~\ref{coro:goodcode}.
Note that $\rvxiest=\rvxi_{\rvt'}$,
where $\rvt'\subset\{ i \mid \rva_i=\rvb_i=0 \}$ stands for
the places for the estimation,
and $\rvzetaest=\rvzeta_{\rvt_1}$ with $\rvt_1=\{ i \mid \rva_i=\rvb_i=1 \}$.

Let $\Fcnd{k,\rlss}$ be 
$\crd{\cS_n}^{-1}\sum_{\pi\in\cS_n}P_{[\rvxicode, \rvzetacode]|\rvk=k,\rvss=\rlss} \{ J[\prm(\Gamma+C)] \}$,
where $\rvxicode=\rvxi_{\rvt}$
and $\rvzetacode=\rvzeta_{\rvt}$.
Then, 
\begin{equation}\label{eq:unionbound}
1-\Fcnd{k,\rlss}\le
\Bprv{\rvxicode|\rvk=k,\rvss=\rlss}+
\Bprv{\rvzetacode|\rvk=k,\rvss=\rlss},
\end{equation}
where $B(Q)=\crd{\cS_n}^{-1}\sum_{\pi\in\cS_n} Q[\pi(\Gamma+C)\cmple]$.
The part bounding $B(p)$ in Section~\ref{sec:exp} (the last paragraph in Section~\ref{ss:exp3}), as well as its modification for $\dmn=2$ in Section~\ref{ss:exp2},
applies
verbatim to the present case, 
where we want to upper-bound $\Bprv{\rvgn|\rvk=k,\rvss=(T_{\rm sift},M,n)}$
for $\rvgn=\rvxicode,\rvzetacode$,
if we replace $p^n$, $B(p)$ in (\ref{eq:pr0})
and $\dmn^{-nD(p||Q)}$ in (\ref{eq:lastchain})
by $P_{\rvgn|\rvk=k,\rvss=\rlss,\rvsss=\rlsss}$,
$\Bprv{\rvgn|\rvk=k,\rvss=\rlss,\rvsss=\rlsss}$
and $P_{\rvgn|\rvk=k,\rvss=\rlss,\rvsss=\rlsss}(\cT_Q^n)$, respectively.
Thus, we have
\begin{eqnarray*}
\lefteqn{\Bprv{\rvxicode|\rvk=k,\rvss=\rlss,\rvsss=\rlsss}} \\ 
&\le & \crd{\cP_n}\wkb \sum_{Q\in\cP_n} \Pxncnd (\cT_Q^n)\\
&&\mbox{} \cdot \min\{ \sum_{Q':\, H(Q') \le H(Q)} \dmn^{nH(Q') -\frac{n-k}{2}-1}, 1 \}\\
&\le & \dmn^{\dmn}\crd{\cP_n}^2 \sum_{Q\in\cP_n} 
\Pr\{ \sP_{\rvxicode}=Q | \rvk=k,\rvss=\rlss,\rvsss=\rlsss \}\\
& &  \cdot \dmn^{- n |1-R(n,\rlms,\rlestA,\rlestB)-2H(Q)|^+/2}
\end{eqnarray*}
for any $k$ and $\rlsss$ with
$\Pr\{ \rvk=k,\rvsss=\rlsss | \rvss=\rlss \}>0$,
as well as the counterpart for $\rvzetacode$.
Substituting (\ref{eq:chosenrate}) into these estimates,
applying the operation
$A \mapsto \sum_{\rlsss,k} \Pr\{\rvsss=\rlsss,\rvk=k|\rvss=\rlss\} A$, 
and combining them with (\ref{eq:unionbound}),
we have the following bound on
$F_{\rlss}
=\sum_{k=0}^{n} \Pr\{\rvk=k|\rvss=\rlss \} F_{k,\rlss}$:
\begin{eqnarray}
\lefteqn{1 - F_{\rlss}} \nonumber\\
&\le &
\dmn^{\dmn}\crd{\cP_n}^{2}
\sum_{Q\in\cP_n,Q'\in\cP_{(\rlms-n)/2}} 
\Pr \{ \sP_{\rvxicode}=Q 
\mbox{ and } \sP_{\rvxiest}=Q' |\rvss=\rlss \} \nonumber\\
&&\mbox{}\ \ \ \cdot\dmn^{- n [ |H(Q') - H(Q) + \gamma|^+ ]}\nonumber\\
&&\mbox{}+\dmn^{\dmn}\crd{\cP_n}^{2}\sum_{Q\in\cP_n,Q'\in\cP_{(\rlms-n)/2}}
\Pr \{ \sP_{\rvzetacode}=Q 
\mbox{ and } \sP_{\rvzetaest}=Q' |\rvss=\rlss \} \nonumber\\
&&\mbox{}\ \ \ \cdot\dmn^{- n [ |H(Q') - H(Q) + \gamma|^+ ]}\nonumber\\
&\le &
4\dmn^{\dmn}\crd{\cP_n}^{3} \crd{\cP_{(n+\rlms)/2}}^{3}\dmn^{-n
\Ejoint(\gamma,\alpha)}, \label{eq:joint1}
\end{eqnarray}
where $\alpha=(\rlms-n)/(\rlms+n)$, 
\[
\Ejoint(\gamma,\alpha)
=\min_{0\le \vep \le 2} 
\{ (1-\alpha)^{-1}[g(\alpha)\vep]^2/\Kd+|\gamma-\theta(\vep)|^+ \},
\]
\[
\theta(x)= \left\{ 
\begin{array}{lll}
0 & \mbox{for} & x=0 \\
-x\log_{\dmn}(x/\dmn) & \mbox{for} & 0< x \le 1/2\\ 
1 & \mbox{for} & 1/2<x,
\end{array}
\right.
\]
and
\[
g(\alpha) = \frac{\sqrt{\alpha(1-\alpha)}}{\sqrt{\alpha}+\sqrt{1-\alpha}}.
\]
To see the last inequality in (\ref{eq:joint1}), 
we need the next lemma with $\cY=\myF$ and $N=(\rlms+n)/2$
as well as the continuity of entropy, i.e., that
$\| Q-Q' \|_1 \le \vep$ implies 
$|H(Q)-H(Q')| \le \theta(\vep)$~\cite{CsiszarKoerner};
we have upper-bounded each summation on the right-hand side
by $\sum_{\vep} 2 \crd{\cP_{N}}^{2} \dmn^{-N[g(\alpha)\vep]^2/\Kd}
\dmn^{-n |\gamma-\theta(\vep)|^+ }$ using the lemma, 
where $\vep$ ranges over 
$\{ \vep \mid \exists Q\in\cP_n,Q'\in\cP_{(\rlms-n)/2},\ \| Q-Q' \|_1 = \vep \}$,
which is not greater than
$2 \crd{\cP_n} \crd{\cP_{N}}^{3} \dmn^{-n \Ejoint(\gamma,\alpha)}$.

\begin{lemma}[Random Sampling] \label{lem:sampling}
Let a finite alphabet $\cY$ and positive integers $n$ and $N$,
$0<n<N$, be given.
Put $\rvrb=(N-n)/N$.
Assume that $\rvy$ is an arbitrary random variable taking values in $\cY^{N}$
and we choose $n$ symbols from $\rvy$ uniformly randomly.
Denote the resulting string by $\rvy'$ (arranged in an arbitrary order, 
which does not matter) 
and the string of the remaining digits by $\rvy''$.
Then the probability that
$\|\sP_{\rvy'}-\sP_{\rvy''}\|_1 \ge \vep$ is upper-bounded by
$2 |\cP_N(\cY)|^2 \dmn^{-N[g(\alpha)\vep]^2/\Kd}$.
 \end{lemma}

{\em Proof}\/.
For a fixed realization $y$ of $\rvy$, denote the conditional probability
$\Pr\{ \mbox{$\sP_{\rvy'}=Q$ and $\sP_{\rvy''}=Q'$} | \rvy=y\}$
by $W(Q,Q'|y)$
($W$ is a classical channel).
For now imagine that
$\rvy$ is the sequence of independent random variables identically distributed
according to $Q\in\cP_N(\cY)$, and 
let $(Q^N \times W)[A]$, or the $Q^N \times W$-probability of $A$,
denote the probability of the event $A$ under this condition.
The $Q^N \times W$-probability 
that $\|\sP_{\rvy'}-\sP_{\rvy}\|_1 \ge \vep'/\sqrt{1-\alpha}$ or 
$\|\sP_{\rvy''}-\sP_{\rvy}\|_1 \ge \vep'/\sqrt{\alpha}$ is
upper-bounded by
$2 |\cP_N(\cY)| \dmn^{-N \vep'^2/\Kd}$
by large deviation theory, 
i.e., by (\ref{eq:prob_type}),
and Pinsker's inequality 
$D(Q||Q') \ge \|Q-Q'\|_1^2/\Kd$~\cite{CsiszarKoerner}.
In words, the probability that
$\|\sP_{\rvy'}-\sP_{\rvy}\|_1 < \vep'/\sqrt{1-\alpha}$ and
$\|\sP_{\rvy''}-\sP_{\rvy}\|_1 < \vep'/\sqrt{\alpha}$ is
lower-bounded by
$1-2 |\cP_N(\cY)| \dmn^{-N \vep'^2/\Kd}$.
By the triangle inequality, this immediately implies
$(Q^N\times W)[
\|\sP_{\rvy'}-\sP_{\rvy''}\|_1 < (1/\sqrt{\alpha}+1/\sqrt{1-\alpha})\vep']
\ge 1-2 |\cP_N(\cY)| \dmn^{-N \vep'^2/\Kd}$.
Note that $W(\cdot,\cdot|y)$ is the same for all $y\in\cY^N$ of a fixed type,
and hence, $\forall Q',Q''$, $W(Q',Q''|y) \le (Q^N \times W)[\sP_{\rvy'}= Q' \mbox{ and } \sP_{\rvy''}=Q'' ] / Q^N(\cT_{Q}^{N})$,
where $Q=\sP_y$, for any $y$.
Since for any $Q\in \cP_{N}(\cY)$, $Q^N(\cT_{Q}^{N}) \ge \crd{\cP_{N}(\cY)}^{-1}$ 
(in fact, $\max_{P\in \cP_N(\cY)} Q^N(\cT^{N}_{P})= Q^N(\cT^{N}_{Q})$~\cite{CsiszarKoerner}),
we have
$\Pr\{ \|\sP_{\rvy'}-\sP_{\rvy''}\|_1 \ge (1/\sqrt{\alpha}+1/\sqrt{1-\alpha})\vep' | \rvy=y \}$
$ \le 2 |\cP_N(\cY)|^2 \dmn^{-N \vep'^2/\Kd}$.
Noticing this bound is independent of $y$, we obtain the lemma.
\enproof

{\em Remark}.\/ In the above application of this lemma, 
$\rvy'$ and $\rvy''$ are the code digits and estimation digits,
respectively. This ensures that
$\sP_{\rvy'}$ and $\sP_{\rvy''}$ are close with 
high probability.
In the binary case where $\cY=\{ 0,1 \}$,
upper bounds of the form $\exp\{ -(N-n) K \vep^2 \}$,
with some constant $K$,
for the probability that $\sP_{y}(1)-\sP_{\rvy''}(1)=\vep$ 
has been known for long~\cite{hoeffding63} and most security proofs
for QKD use this type of bounds (e.g., \cite[Appendix~M, e-Print]{biham99}, \cite[Lemma 1]{LoChauA00}, \cite[Eq.~(25)]{GottesmanPreskill01}, \cite[Appendix, Property 16]{InamoriLutkenhausMayers01}, \cite[p.~589, Exercise~12.27]{NielsenChuang}, \cite{hayashi03pc}).
An advantage of the above lemma is the applicability to the case where 
$\crd{\cY}>2$.

The rest of the task is to relate the fidelity bound in (\ref{eq:joint1})
to the mutual information as we did in Section~\ref{sec:security} for individual attacks.
In the present case, we initially have
\begin{eqnarray*}
\lefteqn{I(\mbm{\sigma};\rve|\rvx'\rvz'\rvestA\rvestB\rvpi\rvt\rva\rvb\rvcode\rvk,\rvss=\rlss)}\\
&\le& 2 \dmn^{- n \Ejoint(\gamma,\alpha) +o_1(n,\rlms)} [n(\Ejoint(\gamma,\alpha)+1)-o_1(n,\rlms)]
\end{eqnarray*}
with a negligible function $o_1(n,\rlms)$.
Note that $\mbm{\sigma}$ is independent of
$\rvx',\rvz',\rvestA,\rvestB,\rvpi,\rvt,\rva,\rvb$ and $\rvcode$
conditionally on $\rvk$ (i.e., $\rvssss=(\rvx',\rvz',\rvestA,\rvestB,\rvpi,\rvt, \rva,\rvb,\rvcode)$, $\rvk$ and $\mbm{\sigma}$ form a Markov chain in this order~\cite{CsiszarKoerner})
given  $\rvss=\rlss$ 
since 
the probability of $\mbm{\sigma}$ conditioned on
$\rvk=k$, $\rvssss=\rlssss$ and 
$\rvss=\rlss$ is uniform over $\myFpower{k}$.
By the chain rule
for mutual information, 
again, 
this implies
\[
I(\mbm{\sigma};\rve\rvx'\rvestA\rvestB\rvpi\rvt\rva\rvb\rvcode|\rvk,\rvss=\rlss)
\le \dmn^{- n\Ejoint(\gamma,\alpha) +o(m)}
\]
[cf.\ (\ref{eq:Sch7})],
where $o(m)$ can be explicitly given as
$3\log_{\dmn} 2 + \dmn + 6(\dmn-1) \log_{\dmn} m  + \log_{\dmn}[m(\gamma+1)]$. 
This simultaneously upper-bounds 
$\Pr\{ \mbm{\sigma} \ne \mbm{\sigma}' | \rvss=\rlss \}$
since the argument in Section~A.3 
also extends to the present case trivially.
The bound is valid for $m$ finite
and is also meaningful in the limit of $m$ large
since $\alpha$ goes to $\vra$ in (\ref{eq:code2whole}) almost surely
by the law of large numbers applied to
the stochastic process $\{ (\rva_i,\rvb_i) \}_i$.
In fact, for the almost sure event where
$\alpha \in [r_0,r_1]$ for all large enough $m$,
where $r_0 < r <r_1$, the bound is true with
$\Ejoint(\gamma,\alpha)$ replaced by 
\[
E_2(\gamma,r_0,r_1)= \min_{0 \le \vep \le 2}
[G\vep^2/\Kd + |\gamma-\theta(\vep)|^+].
\]
Here, $G=\min_{r_0 \le \alpha\le r_1} (1-\alpha)^{-1} [g(\alpha)]^2$ can be made positive
so that $E_2(\gamma,r_0,r_1)$ is positive
by choosing $r_0$, $r_1$ and $\gamma$ appropriately.

This protocol achieves the rate 
$(1-\pa-\pb)[1-2\max\{H(\bar{P_{\cA}}),H(\dbar{P_{\cA}})\}]$
for an individual attack $\cA$, 
as can be checked by modifying the argument in Section~\ref{sec:security} more easily.

\mysectionapp{Nomenclature \label{app:nom}}

Several symbols often used in this paper are listed below.

\subsection*{Strings, Probability Distributions and the Weyl Unitary Basis}
\begin{itemize}
\item $\zrv=(0,\dots,0)\in\myFpower{n}$, $1^n=(1,\dots,1)\in\myFpower{n}$
\item $\cX=\myFpower{2}=\myF\times\myF$ 

\item $[u,w]=\big((u_1,w_1),\dots,(u_n,w_n)\big) \in \cX^n$ 
for $u=(u_1,\dots,u_n), w=(w_1,\dots,w_n) \in\myFpower{n}$

\item $\Ebe_{[u,w]}=X^uZ^w$, where $X^u=X^{u_1}\tnsr \cdots \tnsr X^{u_n}$ and
$Z^w=Z^{w_1}\tnsr \cdots \tnsr Z^{w_n}$ 

\item $\sP_y$: type of string $y$, defined by (\ref{eq:type})

\item $\cP(\cY)$: the set of all probability distribution on $\cY$

\item $\cP_n(\cY)$: the set of all types of sequences in $\cY^n$\
[ $\cP_n(\cY)\subset \cP(\cY)$ ]

\item 
$[PQ](x,y)=P(x)Q(y)$ 

\item
$\bar{Q}(\vari)=\sum_{\varj\in\cY} Q(\vari,\varj)$,\
$\dbar{Q}(\vari)=\sum_{\varj\in\cY} Q(\varj,\vari)$

\item $s_T$:
subsequence $s_{j_1}\cdots s_{j_n}$ of $s_1\cdots s_m$,
where $T=\{j_1 \cdots, j_n \}\subset \{ 1, \dots, m \}$ and $j_1<\cdots<j_n$.

\end{itemize}

\subsection*{Standard Notation in Information Theory \label{appsub:IT}}

\begin{itemize}
\item Entropy: $H(P)= - \sum_{y\in\cY} P(y) \log_{\dmn} P(y) $ 
\item Kullback-Leibler information: $D(P||Q)=\sum_{y\in\cY} P(y) \log_{\dmn} \frac{P(y)}{Q(y)}$
\item Mutual information:
For random variables $\sX$ and $\sY$,
$I(\sX;\sY)=D(P_{\sX\sY}||P_{\sX}P_{\sY})$, 
where $P_{\sW}$ denotes the probability distribution of 
$\sW$ for an arbitrary discrete random variable $\sW$;
$I(\sX;\sY|\sZ=z)=D(P_{\sX\sY|\sZ=z}||P_{\sX|\sZ=z}P_{\sY|\sZ=z})$,
where the probability 
that $\sW=w$ conditional on the event $\sZ=z$ is 
denoted by $P_{\sW|\sZ=z}(w)$, 
and $I(\sX;\sY|\sZ)$ stands for the expectation 
$\sum_{z}P_{\sZ}(z)I(\sX;\sY|\sZ=z)$. 
\item $\hmo(x)=-x\log_2 x -(1-x) \log_2 (1-x)$, $0\le x \le 1$
\end{itemize}

\subsection*{CSS Codes} 

\begin{itemize}
\item $\crJ$: transversal 
(set of coset representatives
in which each coset has exactly one representative) 
of $\myFpower{n}/C^{\perp}$

\item $\css{C}{\crJ}$: $\Ebe_{\Kof{\crJ}}$-correcting CSS code made from
a self-orthogonal $C$ with basis $g_1,\dots,g_{\kcl}$, 
where $\Kof{\crJ}$ is given in (\ref{eq:K})

\item Letters $v,x,z$ as coset representatives (after \cite{ShorPreskill00}):\\
$v+\Csm\in \Cbg/\Csm$,\\
$x+\Cbg \in \myFpower{n}/\Cbg$, $z+\Cbg \in \myFpower{n}/\Cbg$
\end{itemize}

\subsection*{Parameters in the BB84 protocol}

\begin{itemize} 
\item $m$: total number of $\dmn$-ary digits transmitted in the BB84 protocol

\item $n$: code-length of CSS code

\item $\kcl=\dim_{\myF} C$

\item $k=n-2\kcl= \log_{\dmn} \dim_{\bC} \cQ_{xz}$
($\cQ_{xz}$: quantum CSS codes)

\end{itemize}

\end{document}